\documentclass[lettersize,onecolumn]{IEEEtran} 

\usepackage{cite}
\usepackage{graphicx}
\usepackage{amssymb}
\usepackage{amsmath}
\usepackage{amsthm}

\usepackage{paralist}
\usepackage{setspace}
\usepackage{microtype}
\usepackage{hyperref}
\usepackage{url}
\usepackage{balance} 
\usepackage{subfigure}
\usepackage{xcolor} 
\usepackage{fancyref}
\usepackage{framed}
\usepackage{booktabs}
\usepackage{multirow}

\usepackage{algorithm}
\usepackage{algorithmic} 
\usepackage{cancel}

\usepackage{amssymb}
\usepackage{amsfonts}
\usepackage{mathrsfs}
\usepackage{xspace}
\usepackage{bm}
\usepackage{upgreek}

\newcommand{\safemath}[2]{\newcommand{#1}{\ensuremath{#2}\xspace}}

\safemath{\bma}{\mathbf{a}}
\safemath{\bmb}{\mathbf{b}}
\safemath{\bmc}{\mathbf{c}}
\safemath{\bmd}{\mathbf{d}}
\safemath{\bme}{\mathbf{e}}
\safemath{\bmf}{\mathbf{f}}
\safemath{\bmg}{\mathbf{g}}
\safemath{\bmh}{\mathbf{h}}
\safemath{\bmi}{\mathbf{i}}
\safemath{\bmj}{\mathbf{j}}
\safemath{\bmk}{\mathbf{k}}
\safemath{\bml}{\mathbf{l}}
\safemath{\bmm}{\mathbf{m}}
\safemath{\bmn}{\mathbf{n}}
\safemath{\bmo}{\mathbf{o}}
\safemath{\bmp}{\mathbf{p}}
\safemath{\bmq}{\mathbf{q}}
\safemath{\bmr}{\mathbf{r}}
\safemath{\bms}{\mathbf{s}}
\safemath{\bmt}{\mathbf{t}}
\safemath{\bmu}{\mathbf{u}}
\safemath{\bmv}{\mathbf{v}}
\safemath{\bmw}{\mathbf{w}}
\safemath{\bmx}{\mathbf{x}}
\safemath{\bmy}{\mathbf{y}}
\safemath{\bmz}{\mathbf{z}}
\safemath{\bmzero}{\mathbf{0}}
\safemath{\bmone}{\mathbf{1}}

\bmdefine{\biad}{a}
\bmdefine{\bibd}{b}
\bmdefine{\bicd}{c}
\bmdefine{\bidd}{d}
\bmdefine{\bied}{e}
\bmdefine{\bifd}{f}
\bmdefine{\bigd}{g}
\bmdefine{\bihd}{h}
\bmdefine{\biid}{i}
\bmdefine{\bijd}{j}
\bmdefine{\bikd}{k}
\bmdefine{\bild}{l}
\bmdefine{\bimd}{m}
\bmdefine{\bind}{n}
\bmdefine{\biod}{o}
\bmdefine{\bipd}{p}
\bmdefine{\biqd}{q}
\bmdefine{\bird}{r}
\bmdefine{\bisd}{s}
\bmdefine{\bitd}{t}
\bmdefine{\biud}{u}
\bmdefine{\bivd}{v}
\bmdefine{\biwd}{w}
\bmdefine{\bixd}{x}
\bmdefine{\biyd}{y}
\bmdefine{\bizd}{z}

\bmdefine{\bixid}{\xi}
\bmdefine{\bilambdad}{\lambda}
\bmdefine{\bimud}{\mu}
\bmdefine{\bithetad}{\theta}
\bmdefine{\biphid}{\phi}
\bmdefine{\bideltad}{\delta}

\safemath{\bmia}{\biad}
\safemath{\bmib}{\bibd}
\safemath{\bmic}{\bicd}
\safemath{\bmid}{\bidd}
\safemath{\bmie}{\bied}
\safemath{\bmif}{\bifd}
\safemath{\bmig}{\bigd}
\safemath{\bmih}{\bihd}
\safemath{\bmii}{\biid}
\safemath{\bmij}{\bijd}
\safemath{\bmik}{\bikd}
\safemath{\bmil}{\bild}
\safemath{\bmim}{\bimd}
\safemath{\bmin}{\bind}
\safemath{\bmio}{\biod}
\safemath{\bmip}{\bipd}
\safemath{\bmiq}{\biqd}
\safemath{\bmir}{\bird}
\safemath{\bmis}{\bisd}
\safemath{\bmit}{\bitd}
\safemath{\bmiu}{\biud}
\safemath{\bmiv}{\bivd}
\safemath{\bmiw}{\biwd}
\safemath{\bmix}{\bixd}
\safemath{\bmiy}{\biyd}
\safemath{\bmiz}{\bizd}

\safemath{\bmxi}{\bixid}
\safemath{\bmlambda}{\bilambdad}
\safemath{\bmmu}{\bimud}
\safemath{\bmtheta}{\bithetad}
\safemath{\bmphi}{\biphid}
\safemath{\bmdelta}{\bideltad}

\safemath{\bA}{\mathbf{A}}
\safemath{\bB}{\mathbf{B}}
\safemath{\bC}{\mathbf{C}}
\safemath{\bD}{\mathbf{D}}
\safemath{\bE}{\mathbf{E}}
\safemath{\bF}{\mathbf{F}}
\safemath{\bG}{\mathbf{G}}
\safemath{\bH}{\mathbf{H}}
\safemath{\bI}{\mathbf{I}}
\safemath{\bJ}{\mathbf{J}}
\safemath{\bK}{\mathbf{K}}
\safemath{\bL}{\mathbf{L}}
\safemath{\bM}{\mathbf{M}}
\safemath{\bN}{\mathbf{N}}
\safemath{\bO}{\mathbf{O}}
\safemath{\bP}{\mathbf{P}}
\safemath{\bQ}{\mathbf{Q}}
\safemath{\bR}{\mathbf{R}}
\safemath{\bS}{\mathbf{S}}
\safemath{\bT}{\mathbf{T}}
\safemath{\bU}{\mathbf{U}}
\safemath{\bV}{\mathbf{V}}
\safemath{\bW}{\mathbf{W}}
\safemath{\bX}{\mathbf{X}}
\safemath{\bY}{\mathbf{Y}}
\safemath{\bZ}{\mathbf{Z}}

\safemath{\bZero}{\mathbf{0}}
\safemath{\bOne}{\mathbf{1}}
\safemath{\bDelta}{\mathbf{\Delta}}
\safemath{\bLambda}{\mathbf{\UpLambda}}
\safemath{\bPhi}{\mathbf{\Upphi}}
\safemath{\bSigma}{\mathbf{\Upsigma}}
\safemath{\bOmega}{\mathbf{\Upomega}}
\safemath{\bTheta}{\mathbf{\Uptheta}}

\bmdefine{\biAd}{A}
\bmdefine{\biBd}{B}
\bmdefine{\biCd}{C}
\bmdefine{\biDd}{D}
\bmdefine{\biEd}{E}
\bmdefine{\biFd}{F}
\bmdefine{\biGd}{G}
\bmdefine{\biHd}{H}
\bmdefine{\biId}{I}
\bmdefine{\biJd}{J}
\bmdefine{\biKd}{K}
\bmdefine{\biLd}{L}
\bmdefine{\biMd}{M}
\bmdefine{\biOd}{N}
\bmdefine{\biPd}{O}
\bmdefine{\biQd}{P}
\bmdefine{\biRd}{R}
\bmdefine{\biSd}{S}
\bmdefine{\biTd}{T}
\bmdefine{\biUd}{U}
\bmdefine{\biVd}{V}
\bmdefine{\biWd}{W}
\bmdefine{\biXd}{X}
\bmdefine{\biYd}{Y}
\bmdefine{\biZd}{Z}

\bmdefine{\biDelta}{\Delta}
\bmdefine{\biLambda}{\Lambda}
\bmdefine{\biPhi}{\Phi}
\bmdefine{\biSigma}{\Sigma}
\bmdefine{\biOmega}{\Omega}
\bmdefine{\biTheta}{\Theta}

\safemath{\bimA}{\biAd}
\safemath{\bimB}{\biBd}
\safemath{\bimC}{\biCd}
\safemath{\bimD}{\biDd}
\safemath{\bimE}{\biEd}
\safemath{\bimF}{\biFd}
\safemath{\bimG}{\biGd}
\safemath{\bimH}{\biHd}
\safemath{\bimI}{\biId}
\safemath{\bimJ}{\biJd}
\safemath{\bimK}{\biKd}
\safemath{\bimL}{\biLd}
\safemath{\bimM}{\biMd}
\safemath{\bimN}{\biNd}
\safemath{\bimO}{\biOd}
\safemath{\bimP}{\biPd}
\safemath{\bimQ}{\biQd}
\safemath{\bimR}{\biRd}
\safemath{\bimS}{\biSd}
\safemath{\bimT}{\biTd}
\safemath{\bimU}{\biUd}
\safemath{\bimV}{\biVd}
\safemath{\bimW}{\biWd}
\safemath{\bimX}{\biXd}
\safemath{\bimY}{\biYd}
\safemath{\bimZ}{\biZd}

\safemath{\bimDelta}{\biDelta}
\safemath{\bimLambda}{\biLambda}
\safemath{\bimPhi}{\biPhi}
\safemath{\bimSigma}{\biSigma}
\safemath{\bimOmega}{\biOmega}
\safemath{\bimTheta}{\biTheta}

\safemath{\setA}{\mathcal{A}}
\safemath{\setB}{\mathcal{B}}
\safemath{\setC}{\mathcal{C}}
\safemath{\setD}{\mathcal{D}}
\safemath{\setE}{\mathcal{E}}
\safemath{\setF}{\mathcal{F}}
\safemath{\setG}{\mathcal{G}}
\safemath{\setH}{\mathcal{H}}
\safemath{\setI}{\mathcal{I}}
\safemath{\setJ}{\mathcal{J}}
\safemath{\setK}{\mathcal{K}}
\safemath{\setL}{\mathcal{L}}
\safemath{\setM}{\mathcal{M}}
\safemath{\setN}{\mathcal{N}}
\safemath{\setO}{\mathcal{O}}
\safemath{\setP}{\mathcal{P}}
\safemath{\setQ}{\mathcal{Q}}
\safemath{\setR}{\mathcal{R}}
\safemath{\setS}{\mathcal{S}}
\safemath{\setT}{\mathcal{T}}
\safemath{\setU}{\mathcal{U}}
\safemath{\setV}{\mathcal{V}}
\safemath{\setW}{\mathcal{W}}
\safemath{\setX}{\mathcal{X}}
\safemath{\setY}{\mathcal{Y}}
\safemath{\setZ}{\mathcal{Z}}
\safemath{\emptySet}{\varnothing}

\safemath{\colA}{\mathscr{A}}
\safemath{\colB}{\mathscr{B}}
\safemath{\colC}{\mathscr{C}}
\safemath{\colD}{\mathscr{D}}
\safemath{\colE}{\mathscr{E}}
\safemath{\colF}{\mathscr{F}}
\safemath{\colG}{\mathscr{G}}
\safemath{\colH}{\mathscr{H}}
\safemath{\colI}{\mathscr{I}}
\safemath{\colJ}{\mathscr{J}}
\safemath{\colK}{\mathscr{K}}
\safemath{\colL}{\mathscr{L}}
\safemath{\colM}{\mathscr{M}}
\safemath{\colN}{\mathscr{N}}
\safemath{\colO}{\mathscr{O}}
\safemath{\colP}{\mathscr{P}}
\safemath{\colQ}{\mathscr{Q}}
\safemath{\colR}{\mathscr{R}}
\safemath{\colS}{\mathscr{S}}
\safemath{\colT}{\mathscr{T}}
\safemath{\colU}{\mathscr{U}}
\safemath{\colV}{\mathscr{V}}
\safemath{\colW}{\mathscr{W}}
\safemath{\colX}{\mathscr{X}}
\safemath{\colY}{\mathscr{Y}}
\safemath{\colZ}{\mathscr{Z}}

\safemath{\opA}{\mathbb{A}}
\safemath{\opB}{\mathbb{B}}
\safemath{\opC}{\mathbb{C}}
\safemath{\opD}{\mathbb{D}}
\safemath{\opE}{\mathbb{E}}
\safemath{\opF}{\mathbb{F}}
\safemath{\opG}{\mathbb{G}}
\safemath{\opH}{\mathbb{H}}
\safemath{\opI}{\mathbb{I}}
\safemath{\opJ}{\mathbb{J}}
\safemath{\opK}{\mathbb{K}}
\safemath{\opL}{\mathbb{L}}
\safemath{\opM}{\mathbb{M}}
\safemath{\opN}{\mathbb{N}}
\safemath{\opO}{\mathbb{O}}
\safemath{\opP}{\mathbb{P}}
\safemath{\opQ}{\mathbb{Q}}
\safemath{\opR}{\mathbb{R}}
\safemath{\opS}{\mathbb{S}}
\safemath{\opT}{\mathbb{T}}
\safemath{\opU}{\mathbb{U}}
\safemath{\opV}{\mathbb{V}}
\safemath{\opW}{\mathbb{W}}
\safemath{\opX}{\mathbb{X}}
\safemath{\opY}{\mathbb{Y}}
\safemath{\opZ}{\mathbb{Z}}
\safemath{\opZero}{\mathbb{O}}
\safemath{\identityop}{\opI}

\safemath{\veca}{\bma}
\safemath{\vecb}{\bmb}
\safemath{\vecc}{\bmc}
\safemath{\vecd}{\bmd}
\safemath{\vece}{\bme}
\safemath{\vecf}{\bmf}
\safemath{\vecg}{\bmg}
\safemath{\vech}{\bmh}
\safemath{\veci}{\bmi}
\safemath{\vecj}{\bmj}
\safemath{\veck}{\bmk}
\safemath{\vecl}{\bml}
\safemath{\vecm}{\bmm}
\safemath{\vecn}{\bmn}
\safemath{\veco}{\bmo}
\safemath{\vecp}{\bmp}
\safemath{\vecq}{\bmq}
\safemath{\vecr}{\bmr}
\safemath{\vecs}{\bms}
\safemath{\vect}{\bmt}
\safemath{\vecu}{\bmu}
\safemath{\vecv}{\bmv}
\safemath{\vecw}{\bmw}
\safemath{\vecx}{\bmx}
\safemath{\vecy}{\bmy}
\safemath{\vecz}{\bmz}

\safemath{\veczero}{\bmzero}
\safemath{\vecone}{\bmone}
\safemath{\vecxi}{\bmxi}
\safemath{\veclambda}{\bmlambda}
\safemath{\vecmu}{\bmmu}
\safemath{\vectheta}{\bmtheta}
\safemath{\vecphi}{\bmphi}
\safemath{\vecdelta}{\bmdelta}

\safemath{\matA}{\bA}
\safemath{\matB}{\bB}
\safemath{\matC}{\bC}
\safemath{\matD}{\bD}
\safemath{\matE}{\bE}
\safemath{\matF}{\bF}
\safemath{\matG}{\bG}
\safemath{\matH}{\bH}
\safemath{\matI}{\bI}
\safemath{\matJ}{\bJ}
\safemath{\matK}{\bK}
\safemath{\matL}{\bL}
\safemath{\matM}{\bM}
\safemath{\matN}{\bN}
\safemath{\matO}{\bO}
\safemath{\matP}{\bP}
\safemath{\matQ}{\bQ}
\safemath{\matR}{\bR}
\safemath{\matS}{\bS}
\safemath{\matT}{\bT}
\safemath{\matU}{\bU}
\safemath{\matV}{\bV}
\safemath{\matW}{\bW}
\safemath{\matX}{\bX}
\safemath{\matY}{\bY}
\safemath{\matZ}{\bZ}
\safemath{\matzero}{\bmzero}

\safemath{\matDelta}{\bDelta}
\safemath{\matLambda}{\bLambda}
\safemath{\matPhi}{\bPhi}
\safemath{\matSigma}{\bSigma}
\safemath{\matOmega}{\bOmega}
\safemath{\matTheta}{\bTheta}

\safemath{\matidentity}{\matI}
\safemath{\matone}{\matO}

\safemath{\rnda}{A}
\safemath{\rndb}{B}
\safemath{\rndc}{C}
\safemath{\rndd}{D}
\safemath{\rnde}{E}
\safemath{\rndf}{F}
\safemath{\rndg}{G}
\safemath{\rndh}{H}
\safemath{\rndi}{I}
\safemath{\rndj}{J}
\safemath{\rndk}{K}
\safemath{\rndl}{L}
\safemath{\rndm}{M}
\safemath{\rndn}{N}
\safemath{\rndo}{O}
\safemath{\rndp}{P}
\safemath{\rndq}{Q}
\safemath{\rndr}{R}
\safemath{\rnds}{S}
\safemath{\rndt}{T}
\safemath{\rndu}{U}
\safemath{\rndv}{V}
\safemath{\rndw}{W}
\safemath{\rndx}{X}
\safemath{\rndy}{Y}
\safemath{\rndz}{Z}

\safemath{\rveca}{\bimA}
\safemath{\rvecb}{\bimB}
\safemath{\rvecc}{\bimC}
\safemath{\rvecd}{\bimD}
\safemath{\rvece}{\bimE}
\safemath{\rvecf}{\bimF}
\safemath{\rvecg}{\bimG}
\safemath{\rvech}{\bimH}
\safemath{\rveci}{\bimI}
\safemath{\rvecj}{\bimJ}
\safemath{\rveck}{\bimK}
\safemath{\rvecl}{\bimL}
\safemath{\rvecm}{\bimM}
\safemath{\rvecn}{\bimN}
\safemath{\rveco}{\bomO}
\safemath{\rvecp}{\bimP}
\safemath{\rvecq}{\bimQ}
\safemath{\rvecr}{\bimR}
\safemath{\rvecs}{\bimS}
\safemath{\rvect}{\bimT}
\safemath{\rvecu}{\bimU}
\safemath{\rvecv}{\bimV}
\safemath{\rvecw}{\bimW}
\safemath{\rvecx}{\bimX}
\safemath{\rvecy}{\bimY}
\safemath{\rvecz}{\bimZ}

\safemath{\rvecxi}{\bmxi}
\safemath{\rveclambda}{\bmlambda}
\safemath{\rvecmu}{\bmmu}
\safemath{\rvectheta}{\bmtheta}
\safemath{\rvecphi}{\bmphi}

\safemath{\rmatA}{\bimA}
\safemath{\rmatB}{\bimB}
\safemath{\rmatC}{\bimC}
\safemath{\rmatD}{\bimD}
\safemath{\rmatE}{\bimE}
\safemath{\rmatF}{\bimF}
\safemath{\rmatG}{\bimG}
\safemath{\rmatH}{\bimH}
\safemath{\rmatI}{\bimI}
\safemath{\rmatJ}{\bimJ}
\safemath{\rmatK}{\bimK}
\safemath{\rmatL}{\bimL}
\safemath{\rmatM}{\bimM}
\safemath{\rmatN}{\bimN}
\safemath{\rmatO}{\bimO}
\safemath{\rmatP}{\bimP}
\safemath{\rmatQ}{\bimQ}
\safemath{\rmatR}{\bimR}
\safemath{\rmatS}{\bimS}
\safemath{\rmatT}{\bimT}
\safemath{\rmatU}{\bimU}
\safemath{\rmatV}{\bimV}
\safemath{\rmatW}{\bimW}
\safemath{\rmatX}{\bimX}
\safemath{\rmatY}{\bimY}
\safemath{\rmatZ}{\bimZ}

\safemath{\rmatDelta}{\bimDelta}
\safemath{\rmatLambda}{\bimLambda}
\safemath{\rmatPhi}{\bimPhi}
\safemath{\rmatSigma}{\bimSigma}
\safemath{\rmatOmega}{\bimOmega}
\safemath{\rmatTheta}{\bimTheta}

\usepackage{amssymb}
\usepackage{amsfonts}
\usepackage{mathrsfs}
\usepackage{xspace}
\usepackage{bm}
\usepackage{fancyref}
\usepackage{textcomp}

\usepackage{multirow}
\usepackage{stmaryrd}

\newenvironment{textbmatrix}{	\setlength{\arraycolsep}{2.5pt}%
								\big[\begin{matrix}}{\end{matrix}\big]%
								\raisebox{0.08ex}{\vphantom{M}}}

\def\be{\begin{equation}}
\def\ee{\end{equation}}
\def\een{\nonumber \end{equation}}
\def\mat{\begin{bmatrix}}
\def\emat{\end{bmatrix}}
\def\btm{\begin{textbmatrix}}
\def\etm{\end{textbmatrix}}

\def\ba#1\ea{\begin{align}#1\end{align}}
\def\bas#1\eas{\begin{align*}#1\end{align*}}
\def\bs#1\es{\begin{split}#1\end{split}} 
\def\bg#1\eg{\begin{gather}#1\end{gather}}
\def\bml#1\eml{\begin{multline}#1\end{multline}}
\def\bi#1\ei{\begin{itemize}#1\end{itemize}}

\newcommand{\lefto}{\mathopen{}\left}

\DeclareMathOperator{\tr}{tr}				
\DeclareMathOperator*{\argmin}{arg\;min}		
\DeclareMathOperator*{\argmax}{arg\;max}		
\DeclareMathOperator{\grad}{\nabla}			

\newcommand{\Ex}[2]{\ensuremath{\Exop_{#1}\lefto[#2\right]}} 	

\newcommand{\vecnorm}[1]{\lefto\lVert#1\right\rVert}		

\safemath{\dirac}{\delta}					
\safemath{\krond}{\dirac}					

\safemath{\upto}{\uparrow}
\safemath{\downto}{\downarrow}
\safemath{\iu}{j}							
\safemath{\ev}{\lambda}						
\safemath{\hilseqspace}{l^{2}}				
\newcommand{\banachfunspace}[1]{\setL^{#1}}	
\safemath{\hilfunspace}{\banachfunspace{2}}	

\safemath{\SNR}{\textsf{SNR}} 				
\safemath{\PAR}{\textsf{PAR}} 				
\safemath{\No}{N_0}							
\safemath{\Es}{E_s}							
\safemath{\Eb}{E_b}							
\safemath{\EbNo}{\frac{\Eb}{\No}}
\safemath{\EsNo}{\frac{\Es}{\No}}

\DeclareMathOperator{\CHop}{\ensuremath{\opH}} 
\safemath{\tvir}{\rndh_{\CHop}}				
\safemath{\tvtf}{\rndl_{\CHop}}				
\safemath{\spf}{\rnds_{\CHop}}				
\safemath{\bff}{H_{\CHop}}					

\safemath{\ircf}{r_{h}}						
\safemath{\tftvcf}{r_{s}}					
\safemath{\tfcf}{r_{l}}						
\safemath{\bfcf}{r_{H}}						

\safemath{\tcorr}{c_h}						
\safemath{\scf}{c_{s}}						
\safemath{\tfcorr}{c_{l}}					
\safemath{\fcorr}{c_{H}}						

\safemath{\mi}{I}							
\safemath{\capacity}{C}						

\safemath{\normal}{\mathcal{N}}			
\safemath{\jpg}{\mathcal{CN}}			
\safemath{\mchain}{\leftrightarrow}		

\safemath{\dB}{\,\mathrm{dB}}
\safemath{\dBm}{\,\mathrm{dBm}}
\safemath{\Hz}{\,\mathrm{Hz}}
\safemath{\kHz}{\,\mathrm{kHz}}
\safemath{\MHz}{\,\mathrm{MHz}}
\safemath{\GHz}{\,\mathrm{GHz}}
\safemath{\s}{\,\mathrm{s}}
\safemath{\ms}{\,\mathrm{ms}}
\safemath{\mus}{\,\mathrm{\text{\textmu}s}}
\safemath{\ns}{\,\mathrm{ns}}
\safemath{\ps}{\,\mathrm{ps}}
\safemath{\meter}{\,\mathrm{m}}
\safemath{\mm}{\,\mathrm{mm}}
\safemath{\cm}{\,\mathrm{cm}}
\safemath{\m}{\,\mathrm{m}}
\safemath{\W}{\,\mathrm{W}}
\safemath{\mW}{\, \mathrm{mW}}
\safemath{\J}{\,\mathrm{J}}
\safemath{\K}{\,\mathrm{K}}
\safemath{\bit}{\,\mathrm{bit}}
\safemath{\nat}{\,\mathrm{nat}}

\safemath{\define}{\triangleq}			

\safemath{\equivalent}{\sim}
\safemath{\distas}{\sim}					
\safemath{\sdiff}{\Delta}				

\safemath{\reals}{\mathbb{R}}
\safemath{\positivereals}{\reals_{+}}
\safemath{\integers}{\mathbb{Z}}
\safemath{\posint}{\integers_{+}}
\safemath{\naturals}{\mathbb{N}}
\safemath{\posnaturals}{\naturals_{+}}
\safemath{\complexset}{\mathbb{C}}
\safemath{\rationals}{\mathbb{Q}}

\newcommand*{\fancyrefapplabelprefix}{app}		
\newcommand*{\fancyrefthmlabelprefix}{thm}		
\newcommand*{\fancyreflemlabelprefix}{lem}		
\newcommand*{\fancyrefcorlabelprefix}{cor}		
\newcommand*{\fancyrefdeflabelprefix}{def}		
\newcommand*{\fancyrefproplabelprefix}{prop}	
\newcommand*{\fancyrefobslabelprefix}{obs}		
\newcommand*{\fancyrefalglabelprefix}{alg}		
\newcommand*{\fancyrefasmlabelprefix}{asm}	    
\newcommand*{\fancyreftbllabelprefix}{tbl}	    
\newcommand*{\fancyrefremlabelprefix}{rem}	    

\frefformat{vario}{\fancyrefseclabelprefix}{Section~#1}
\frefformat{vario}{\fancyrefthmlabelprefix}{Theorem~#1}
\frefformat{vario}{\fancyreflemlabelprefix}{Lemma~#1}
\frefformat{vario}{\fancyrefcorlabelprefix}{Corollary~#1}
\frefformat{vario}{\fancyrefdeflabelprefix}{Definition~#1}
\frefformat{vario}{\fancyrefobslabelprefix}{Observation~#1}
\frefformat{vario}{\fancyrefasmlabelprefix}{Assumption~#1}
\frefformat{vario}{\fancyreffiglabelprefix}{Figure~#1}
\frefformat{vario}{\fancyrefapplabelprefix}{Appendix~#1} 
\frefformat{vario}{\fancyrefproplabelprefix}{Proposition~#1}
\frefformat{vario}{\fancyrefalglabelprefix}{Algorithm~#1}
\frefformat{vario}{\fancyrefeqlabelprefix}{(#1)}
\frefformat{vario}{\fancyreftbllabelprefix}{Table~#1}
\frefformat{vario}{\fancyrefremlabelprefix}{Remark~#1}

\newtheorem{thm}{Theorem}

\newtheorem{lem}[thm]{Lemma} 

\newtheorem{rem}{Remark}

\safemath{\dictab}{[\,\dicta\,\,\dictb\,]}

\safemath{\ysig}{\bmy}
\safemath{\ysighat}{\hat{\ysig}}
\safemath{\ysigdim}{M}
\safemath{\xsig}{\bmx}
\safemath{\xsigdim}{N}
\safemath{\nx}{n_x}
\safemath{\zsig}{\bmz}
\safemath{\zsigdim}{\ysigdim}
\safemath{\rsig}{\bmr}
\safemath{\Adict}{\bA}
\safemath{\Adicttilde}{\widetilde{\Adict}}
\safemath{\Adictdim}{\outputdim\times\xsigdim}
\safemath{\avec}{\bma}
\safemath{\avectilde}{\tilde{\avec}}
\safemath{\Bdict}{\bB}
\safemath{\Bdicttilde}{\widetilde{\Bdict}}
\safemath{\Cdict}{\bC}
\safemath{\cvec}{\bmc}
\safemath{\Ddict}{\bD}
\safemath{\Ddictdim}{\ysigdim\times\xsigdim}
\safemath{\dvec}{\bmd}
\safemath{\Ddicttilde}{\widetilde{\bD}}
\safemath{\Bonb}{\bB}
\safemath{\bvec}{\bmb}
\safemath{\Bonbdim}{\ysigdim\times\ysigdim}
\safemath{\noise}{\bmn}
\safemath{\noisedim}{\ysigim}
\safemath{\err}{\bme}
\safemath{\errdim}{\ysigdim}
\safemath{\errset}{\setE}
\safemath{\nerr}{n_e}
\safemath{\delop}{\bP_\errset}
\safemath{\delopc}{\bP_{{\errset}^c}}

\safemath{\cplxi}{\imath}
\safemath{\cplxj}{\jmath}

\safemath{\dict}{\matD}
\safemath{\inputdim}{N}		
\safemath{\outputdim}{M}		
\safemath{\sparsity}{S}	
\safemath{\inputdimA}{{N_a}}	
\safemath{\inputdimB}{{N_b}}	
\safemath{\elemA}{{n_a}}	
\safemath{\elemB}{{n_b}}	
\safemath{\resA}{\matR_a}	
\safemath{\resB}{\matR_b}	
\safemath{\subD}{\matS} 
\safemath{\subA}{\matS_a} 
\safemath{\subB}{\matS_b} 
\safemath{\dicta}{\matA} 	
\safemath{\dictb}{\matB} 	
\safemath{\hollowS}{H}
\safemath{\hollowA}{H_a}
\safemath{\hollowB}{H_b}
\safemath{\cross}{Z}
\safemath{\coh}{\mu_d}			
\safemath{\coha}{\mu_a}			
\safemath{\cohb}{\mu_b}			
\safemath{\mubs}{\nu}	
\safemath{\cohm}{\mu_m} 
\safemath{\dictset}{\setD}	
\safemath{\dictsetp}{\dictset(\coh,\coha,\cohb)}	
\safemath{\dictsetgen}{\dictset_\text{gen}}
\safemath{\dictsetgenp}{\dictsetgen(\coh)}
\safemath{\dictsetonb}{\dictset_\text{onb}}
\safemath{\dictsetonbp}{\dictsetonb(\coh)}

\safemath{\leftside}{U}
\safemath{\rightsideA}{R_a}
\safemath{\rightsideB}{R_b}

\safemath{\indexS}{\setI_S} 

\safemath{\na}{n_a}			
\safemath{\nb}{n_b}			
\safemath{\coeffa}{p_i}	
\safemath{\coeffb}{q_j}	
\safemath{\seta}{\setP}		
\safemath{\setb}{\setQ}     
\safemath{\setw}{\setW}	
\safemath{\setz}{\setZ}	
\safemath{\cola}{\veca}		
\safemath{\colb}{\vecb}		
\safemath{\cold}{\vecd}		
\safemath{\inputvec}{\vecx} 	
\safemath{\error}{\vece}	
\safemath{\noiseout}{\vecz} 	
\safemath{\inputvecel}{x}
\safemath{\inputveca}{\vecx_a}
\safemath{\inputvecb}{\vecx_b}
\safemath{\outputvec}{\vecy}	
\safemath{\lambdamin}{\lambda_{\mathrm{min}}}

\safemath{\elltwo}{\ell_2}
\safemath{\ellone}{\ell_1}
\safemath{\ellzero}{\ell_0}
\safemath{\ellinf}{\ell_\infty}
\safemath{\ellinftilde}{\ell_{\widetilde\infty}}
\safemath{\licard}{Z(\coh,\coha,\cohb)}
\safemath{\xsol}{\hat{x}}
\safemath{\xbord}{x_b}		
\safemath{\xstat}{x_s}		
\safemath{\xstatLone}{\tilde{x}_s}
\safemath{\order}{\mathcal{O}} 
\safemath{\scales}{\Theta} 
\safemath{\ones}{\mathbf{1}} 
\safemath{\zeroes}{\mathbf{0}} 
\safemath{\thlone}{\kappa(\coh,\cohb)} 
\safemath{\constoneA}{\delta} 
\safemath{\constoneB}{\epsilon} 
\safemath{\nlarge}{L}				   
\safemath{\sumlarge}{S_\nlarge}
\safemath{\maxlarger}{P_\nlarge}	   
\safemath{\Pzero}{\textrm{P0}}	
\safemath{\Pone}{\textrm{P1}}
\safemath{\vecfir}{\vecw}			 
\safemath{\vecsec}{\vecz}
\safemath{\elvecfir}{w}              
\safemath{\elvecsec}{z}				 
\safemath{\nlargefir}{n}
\safemath{\normout}{\gamma}
\safemath{\auxfun}{h}
\safemath{\supp}{\textrm{supp}}

\safemath{\indexa}{\ell}
\safemath{\indexb}{r}
\safemath{\indexc}{i}
\safemath{\indexd}{j}

\safemath{\project}{P}

\newtheorem{alg}{Algorithm} 
\newenvironment{customalg}[1]
  {\renewcommand\thealg{#1}\alg}
  {\endalg}
\newtheorem{theorem}{Theorem}

\newcommand{\half}{\frac{1}{2}}

\allowdisplaybreaks 

\usepackage{color}

\newcommand{\proj}[2]{\mathcal{P}_{#1}\!\left(#2\right)}
\newcommand{\projop}[1]{\mathcal{P}_{#1}}
\def\td{\mathrm{d}}

\newcommand{\gdet}[1]{g\!\left(#1\right)}
\newcommand{\gint}[1]{\hat{g}\!\left(#1\right)}
\newcommand{\SVD}[1]{\textit{SVD}\!\left(#1\right)}
\usepackage{bbm}

\def\bSigma{\mathbf{\Sigma}}
\def\CP{\mathrm{CP}}
\def\Unif{\mathrm{Unif}}
\def\grad{\mathrm{grad \ }}
\newcommand{\pder}[2]{\frac{\partial#1}{\partial#2} }
\safemath{\Tran}{\textnormal{T}}
\safemath{\Herm}{\textnormal{H}}
\def\Fro{\textnormal{F}}
\newcommand{\setint}[1]{[#1]}
\makeatletter 
\newcommand{\leqnomode}{\tagsleft@true}
\newcommand{\reqnomode}{\tagsleft@false}
 
\DeclareUnicodeCharacter{2212}{-}
\DeclareUnicodeCharacter{2032}{'}
\makeatother

\title{Learning Sparsifying Transforms for mmWave Communication via $\ell^4$-Norm Maximization}
\author{\IEEEauthorblockN{Sueda Taner and Christoph Studer}
\thanks{ST is with Ericsson Research in Kista, Sweden and CS is with ETH Zurich in Switzerland; e-mail: sueda.taner@ericsson.com, studer@ethz.ch.}
\thanks{Parts of this paper were published in \cite{taner2021optimality}. 
Here, we provide the theoretical background omitted in \cite{taner2021optimality} through additional theorems and proofs that motivate maximizing the $\ell^4$-norm for learning complete unitary dictionaries; we also include a more detailed analysis of the optimality of the discrete Fourier transform (DFT) for sparsifying channel vectors and a new result for the suboptimality of the discrete cosine transform (DCT) for sparsifying signals from a specific real-valued sinusoidal signal model.}
\thanks{This work was supported in part by the Swiss National Science Foundation (NSF) under grants 218704 and 207314, and by CHIST-ERA grant for the project CHASER (CHIST-ERA-22-WAI-01) through the SNSF grant 218704.}
}

\begin{document}	
	
\maketitle


\begin{abstract}
The high directionality of wave propagation at millimeter-wave (mmWave) carrier frequencies results in only a small number of significant transmission paths between user equipments and the basestation (BS). This sparse nature of wave propagation is revealed in the beamspace domain, which is traditionally obtained by taking the spatial discrete Fourier transform (DFT) across a uniform linear antenna array at the BS, where each DFT output is associated with a distinct beam. In recent years, beamspace processing has emerged as a promising technique to reduce baseband complexity and power consumption in all-digital massive multiuser (MU) multiple-input multiple-output (MIMO) systems operating at mmWave frequencies.
However, it remains unclear whether the DFT is the optimal sparsifying transform for finite-dimensional antenna arrays. 
In this paper, we extend the framework of Zhai~\emph{et al.} for complete dictionary learning via $\ell^4$-norm maximization to the complex case in order to learn new sparsifying transforms. 
We provide a theoretical foundation for $\ell^4$-norm maximization and propose two suitable learning algorithms. We then utilize these algorithms (i) to assess the optimality of the DFT for sparsifying channel vectors theoretically and via simulations and (ii) to learn improved sparsifying transforms for real-world and synthetically generated channel vectors. 
\end{abstract}

%
	

\section{Introduction} 
\label{sec:intro}

\IEEEPARstart{M}{assive} multiuser (MU) multiple-input multiple-output (MIMO) and millimeter-wave (mmWave) communication are key technologies in fifth-generation (5G) and future wireless systems~\cite{larsson14a,rappaport13a}.
In order to counteract the high path loss at mmWave carrier frequencies, beamforming with multi-antenna basestation (BS) architectures is essential.
All-digital BS architectures unlock the full potential of antenna arrays: they enable flexible beamforming, simplify channel estimation, and support spatial multiplexing, while incurring system costs and radio-frequency (RF) power consumption comparable to fully analog or hybrid architectures~\cite{yan2019performance,roth18a,skrimponis20206gsummit}.
However, realizing all-digital BS architectures in practice requires hardware- and power-efficient baseband algorithms for channel estimation, data detection, and MU precoding.

A promising approach towards reducing baseband complexity and simplifying hardware is to exploit beamspace sparsity.
Due to the highly directional nature of mmWave propagation and the resulting strong path loss, only a small number of dominant transmission paths typically exist between the user equipments (UEs) and the BS antenna array~\cite{rappaport13a,rappaport_book}. This property enables sparse representations of mmWave channel vectors.
To obtain such sparse representations, the channel vectors are typically transformed from the antenna domain into the beamspace domain via a spatial discrete Fourier transform (DFT), applied across the antenna array (for a uniform linear array, for example). Each index in the resulting beamspace corresponds to a specific beam direction, yielding large entries in absolute value if a propagation path is present in that direction.

As demonstrated in~\cite{Sayeed13GLOBECOM, schniter14, gao16bs, chen17bscs, ma18,abdelghany2019precoding,mirfarshbafan19a,seyedicassp20,mahdavi20beamspace,mirfarshbafan21tcas,gonultacs2021hardware,osinsky21,dai22jsips,he23twc,alimo24}, exploiting beamspace sparsity enables the design of low-complexity baseband algorithms and hardware architectures for channel estimation, data detection, and MU precoding.
While the use of the DFT for sparsifying channel vectors can be justified theoretically for very large antenna arrays~\cite{sayeed02a,nowak2010}, it remains unclear whether the DFT is the optimal sparsifying transform for realistic, finite-dimensional arrays and for both modeled as well as real-world (e.g., measured) mmWave channel vectors.

\subsection{Contributions}
In this paper, we address this question within the broader framework of learning unitary sparsifying transforms for complex-valued vectors. Our main contributions are as follows:
\begin{itemize}
\item We extend the $\ell^4$-norm maximization framework of~\cite{zhai20complete} from the real-valued orthogonal group to the complex-valued unitary group.
\item We formulate an $\ell^4$-norm-based optimization problem under a complex-valued stochastic data model that enables learning unitary sparsifying transforms for mmWave channels.
\item We adapt the real-valued matching, stretching, and projection (MSP) algorithm of~\cite{zhai20complete} to the proposed complex-valued setting. We characterize the algorithm's fixed points, without determining whether they are local optima or saddle points.
\item We propose an alternative learning algorithm based on coordinate ascent (CA) implemented via a sequence of Givens rotations and complex phase shifts. This algorithm facilitates analysis of the local optimality of its fixed points.
\item We prove that, for a certain multipath channel model, the DFT matrix is a fixed point of the MSP algorithm. Moreover, under a free-space line-of-sight channel model, we demonstrate that the DFT matrix is a fixed point of the CA algorithm and locally maximizes the sparsity-inducing objective.
\item  We establish, through numerical analysis, that the discrete cosine transform (DCT) matrix is not a fixed point of the CA algorithm for a real-valued sinusoidal signal model, revealing that---perhaps surprisingly---the DCT is not an optimal sparsifier for this class of signals.
\item We conduct numerical experiments on both synthetic and real-world channel vectors, comparing sparsity-exploiting baseband algorithm performance when using the DFT versus learned transforms. Our experiments reveal that (i) for synthetic channel data, the DFT is an excellent unitary sparsifier, offering only marginal gains over learned transforms; and (ii) for measured channel vectors from a real-world massive MU-MIMO system, learned transforms significantly improve sparsity, and consequently substantially improve error-rate performance of sparsity-exploiting baseband algorithms.
\end{itemize}

\subsection{Relevant Prior Art}

Learning sparsifying transforms is closely related to dictionary learning~\cite{olshausen1997sparse, KSVD}. Dictionary learning aims at finding a (possibly overcomplete) basis or transform, referred to as a \textit{dictionary}, in which a set of input signals admits a sparse representation.
In wireless communications, dictionary learning algorithms have been utilized for sparse channel estimation~\cite{ding3, huang19,xie20twc,  yu21icicsp, nazzal2022wcnc, zhou22comlet,aygul23,xie23twc,maity23irs,bayraktar24ris}, for reducing the overhead of channel state information feedback~\cite{lu15globecom,gadamsetty23}, and for active user detection~\cite{bai22,liang23}.
These references typically address the following regularized optimization problem (or variants thereof) in which the $\ell^0$-pseudo-norm is replaced by the $\ell^1$-norm for tractability: 
\begin{align}
\underset{\bD\in\opC^{M\times N},\,\bX\in\opC^{N\times T}}{\text{minimize}}\,\, \|\bar{\bH}-\bD\bX\|_{\Fro}^2  + \lambda \|\bX\|_0. \label{eq:regDL}
\end{align}
Here, the matrix $\bar{\bH} \in \opC^{M \times T} $ contains $T$ channel samples, with each column \( \bar{\bmh}_t \in \opC^M \) representing a channel vector from a single transmit antenna to $M$ receiver antennas, the matrix \( \bD \in \opC^{M \times N} \) denotes a (possibly overcomplete) dictionary with $M \leq N$ (one often includes an additional constraint that the columns of $\bD$ have unit length), the matrix \( \bX=[\bmx_1\cdots\, \bmx_T] \in \opC^{N \times T} \) contains the corresponding sparse coefficient vectors $\bmx_t$, $t=1,\ldots,T$, and $\lambda>0$ is a regularization parameter that trades off the error of the sparse representation and the sparsity level.
The goal of this dictionary learning problem is to approximate $\bar{\bH}$ by $\bD\bX$ while minimizing the number of nonzero entries of $\bX$.

An alternative approach to the problem in~\fref{eq:regDL} is to reformulate it as a constrained optimization problem, typically of the form
\begin{align}
\underset{ \bD\in\opC^{M\times N},\, \bX\in\opC^{N\times T}}{\text{minimize}}\,\, \|\bar\bH-\bD\bX\|_{\Fro}^2\,\, \text{ subject to }\,   \|\bmx_t\|_0\leq \eta,\, t=1,\ldots,T.
\end{align}
This problem can, for example, be solved approximately by the widely used K-SVD algorithm~\cite{KSVD} or the method of optimal directions (MOD)~\cite{engan99mod}.
Both of these methods alternate between a \textit{sparse coding} step and a \textit{dictionary update} step:
The sparse coding step finds the new best sparse representation for a fixed dictionary by algorithms such as orthogonal matching pursuit (OMP)~\cite{OMP} or basis pursuit~\cite{chen1998}.
The dictionary update step simultaneously updates the dictionary and the sparse coefficient vectors while the sparsity pattern remains fixed. 

The dictionaries obtained through such approaches in~\cite{ding3, huang19,xie20twc, yu21icicsp, nazzal2022wcnc, zhou22comlet,aygul23,xie23twc,maity23irs,bayraktar24ris,lu15globecom,gadamsetty23,bai22,liang23} are arbitrary matrices (besides unit column norm). Therefore, the sparsifying transforms induced by these dictionaries are, in general, not unitary.
As a result, applying such a transform to the received signal across the BS antenna array would alter the noise statistics, meaning that signal processing tasks performed in the transformed (sparse) domain are no longer statistically equivalent to those in the original antenna domain.
In contrast to such results, we aim to learn \emph{unitary} sparsifying transforms by
building upon a problem setup that uses a smooth, $\ell^4$-norm-based sparsity measure with orthogonal dictionaries as proposed in~\cite{zhai20complete} for real-valued signals.

Reference~\cite{xue20blind} also builds upon the $\ell^4$-norm-based dictionary learning approach in~\cite{zhai20complete},
but for the specific application of blind data detection using $\ell^3$-norm maximization.
Also, in~\cite{liu24ofdm}, the $\ell^4$-norm is used in defining a performance metric for integrated sensing and communication in order to study the optimality of the orthogonal frequency division multiplexing (OFDM) modulation waveform.
In contrast to both of these results, we study the more general problem of \emph{learning} sparsifying transforms for channel vectors via $\ell^4$-norm maximization.

\subsection{Notation}

Bold lowercase and uppercase letters represent column vectors and matrices, respectively. 
For a vector $\bma$, the $k$th entry is denoted by $a_k=[\bma]_{k}$; for a matrix $\bA$, the $k$th column is denoted by $\bma_k$ and the $(i,k)$th entry by $A_{i,k}=[\bA]_{i,k}$. 
The superscripts $(\cdot)^*,(\cdot)^{\Tran}$, and $(\cdot)^{\Herm}$ stand for the matrix conjugate, transpose, and Hermitian, respectively. 
The $N\times M$ all-zeros matrix is $\bZero_{N\times M}$, the $N\times N$ identity matrix is $\bI_N$, and the $N\times N$ unitary DFT matrix is $\bF_N$; the dimensions in the subscripts are omitted when clear from the context.
The sets of $N\times N$ orthogonal matrices and unitary matrices are denoted by $O(N;\opR)$ and $U(N;\opC)$, respectively.
A complex-valued permutation matrix is defined as a unitary matrix in which every row and column has exactly one non-zero entry; the set of complex-valued permutation matrices of dimension $N\times N$ is denoted by $\CP(N)$.
The element-wise multiplication, Kronecker product, absolute value, and $r$th power are denoted by $\circ$, $\otimes$, $|\cdot|$, and $(\cdot)^{\circ r}$, respectively. 
Following~\cite{zhai20complete}, we denote the element-wise $\ell^p$-norm of a matrix~$\bA$ for $p\geq1$ by $\vecnorm{\bA}_p$, so that  $\vecnorm{\bA}_p^p\define\sum_{i,k} |A_{i,k}|^p$.
The \textit{Frobenius norm} of a matrix~$\bA$ is $\|\bA\|_\Fro \define \sqrt{ \sum_{i,k} |A_{i,k}|^2 } $.
The set of $N$-dimensional vectors on the unit sphere (i.e., vectors that have unit $\ell^2$-norm) is $\opS^{N-1}$.
The singular value decomposition of a matrix~$\bA$ is denoted by $\SVD{\bA}$ and the $i$th largest singular value of $\bA$ is denoted by $\sigma_i(\bA)$.
The set of integers $\{1,\dots,N\}$ is $\setint{N}$. 
The real and imaginary parts are indicated by $\Re\{\cdot\}$ and $\Im\{\cdot\}$. 
The expectation operator is $\Ex{}{\cdot}$. 
We use $\delta[n]$ to refer to the Kronecker delta function for $n\in\opZ$, such that $\delta[0]=1$ and $\delta[n]=0$ for $n\neq 0$.
All complex-valued gradients follow the definitions of~\cite{kreutzdelgado2009complex}. 

\subsection{Paper Outline}
The rest of this paper is organized as follows.
\fref{sec:prereqs} introduces the mmWave massive MU-MIMO uplink system and channel models, followed by their beamspace representation. 
\fref{sec:l4setup} motivates the $\ell^4$-norm metric and details the $\ell^4$-norm maximization problem setup. 
\fref{sec:learning} proposes two algorithms to approximately solve the $\ell^4$-norm maximization problem.
\fref{sec:optimality} examines the optimality of the DFT and DCT using these two algorithms.
\fref{sec:results} provides simulation results for the performance of the DFT for sparsifying channel vectors.
\fref{sec:conclusions} concludes. 


\section{System Model}
\label{sec:prereqs}

\subsection{Uplink System Model}
\label{sec:uplink}

We consider a mmWave massive MU-MIMO uplink system in which data transmission is from $U$ single-antenna UEs to a BS equipped with $B$ antennas.
Let $\bH\in\opC^{B\times U}$ represent the uplink channel matrix between the BS and UEs, and $\vecs\in\mathcal{S}^U$ denote the $U$ data symbols taken from a constellation $\mathcal{S}$. 
For narrowband transmission, we express the baseband receive vector in the \emph{antenna domain} as 
\begin{equation} \label{eq:rx_vec}
    \vecy = \bH \vecs + \vecn,
\end{equation}
where $\vecy\in\opC^B$ is the receive vector and $\vecn$ models i.i.d. circularly-symmetric Gaussian noise. 

\subsection{Channel Model} 
\label{sec:channel_model}
We focus on wave propagation at mmWave frequencies \cite{rappaport_book}, and assume a sufficiently large distance between the BS and the UEs (or scatterers). 
Under the assumption of ideal hardware, for a BS equipped with a uniform linear array (ULA) and $\lambda/2$ antenna spacing (with wavelength~$\lambda$), the columns of the MIMO channel matrix~$\bH$ representing the channel between a specific UE and the BS antenna array can be modeled as follows~\cite{tse}:
\begin{align} 
\vech = \sum_{\ell=1}^{L} c_\ell\vecp(\omega_\ell), \quad \vecp(\omega) = \big[ 1,e^{j\omega},\dots,e^{j(B-1)\omega} \big]^{\Tran}. \label{eq:h_sum}
\end{align}
Here, $L\geq 1$ refers to the total number of paths arriving at the antenna array (including a potential line-of-sight path), $c_\ell\in\opC$ is the complex-valued channel gain associated with the $\ell$th propagation path, $\vecp(\omega)$ is the array response vector, and $\omega_\ell$ is the angular frequency usually given by the relation $\omega_\ell=\pi\sin(\phi_\ell)$, where $\phi_\ell$ is the angle of the $\ell$th path relative to the antenna array. 

\begin{rem}
\label{rem:compensation}
In case of hardware nonidealities in the BS antenna array, such as uneven attenuation/phase between antenna elements, malfunctioning antennas, antenna coupling or RF crosstalk, etc., the channel model from \fref{eq:h_sum} would not hold. 
If the hardware impairments are known and can be expressed by an invertible matrix, then the impairments could be compensated by multiplying the inverse of the impairment matrix by the receive vector. 
With such a compensation, the channel input-output relation would reduce to that of \fref{eq:rx_vec}, with the channel model from~\fref{eq:h_sum} again, but the noise statistics would likely be altered; this would require special attention in the subsequent baseband processing operations, e.g., for beamforming or data detection.
\end{rem}

\subsection{Beamspace Representation} 
\label{sec:beamspace}

To reveal the sparse nature of wave propagation at high carrier frequencies, one can obtain the beamspace (or \textit{angular domain}) input-output relation by applying a DFT to the received signal in the BS antenna array \cite{seyedicassp20,mirfarshbafan19a}. 
For a BS equipped with a ULA, this beamspace representation is obtained by $\bar\vecy=\bF_B \vecy$ under the assumption of ideal hardware or known hardware impairments that can be compensated for as discussed in \fref{rem:compensation}.
The beamspace representation transforms the columns of $\bH$, i.e., the superposition of $L$ complex sinusoids in~\fref{eq:h_sum}, into the frequency domain, which results in sparse beamspace channel vectors if $L$ is small. 
However, under unknown or unmitigable impairments, the columns of the channel matrix would no longer be modeled by~\fref{eq:h_sum}, potentially negatively affecting sparsity.
This key fact motivates us to examine the DFT's optimality for maximizing the sparsity of the receive vectors and to \textit{learn} alternative unitary transforms that exhibit superior sparsifying properties to the widely used DFT.

\section{$\ell^4$-Norm as a Measure of Sparsity}
\label{sec:l4setup}

Reference~\cite{zhai20complete} developed an $\ell^4$-norm-based dictionary learning framework over the orthogonal group $O(N;\opR)$ for a set of given real-valued vectors. 
The intuition behind this framework is that maximizing the $\ell^4$-norm of a matrix over a hypersphere promotes sparsity, i.e., the sparsest points on an $\ell^2$-norm-hypersphere have the smallest $\ell^1$-norm and the largest $\ell^4$-norm~\cite{zhai20complete}.
In what follows, we build upon this insight and consider the complex-valued case with a stochastic data model with the goal of using this framework for beamspace processing.

\subsection{Problem Setup}

Suppose that we have data samples that follow a certain stochastic model and we wish to find a unitary transform that maximally sparsifies such data samples in expectation.
Specifically, we measure the expected sparsity after applying a unitary transform $\bA\in U(N;\opC)$ on a data sample    $\vecy\in\opC^N$ with the $\ell^4$-norm-based objective function $\hat{g}: U(N;\opC)\to\opR$  defined as 
\begin{align} \label{eq:gint}
\gint{\bA}&\define \Ex{\bmy}{\vecnorm{\bA\vecy}_4^4} .
\end{align}
We formulate our goal of finding unitary transforms that maximize the sparsity with the following optimization problem:
\begin{align}
 \underset{\bA}{\text{maximize}}\,\, \gint{\bA}  \,\, \text{subject to }\, \bA\in U(N;\opC). \tag{O1}\label{eq:obj1}
\end{align}
We can interpret taking the expectation of the $\ell^4$-norm over the probabilistic distribution of a data sample as averaging the $\ell^4$-norm over an infinitely large set of data samples.
Similarly, averaging the $\ell^4$-norm for a finite set of observation samples is equivalent to taking the expectation over a uniform probability mass function (PMF) that is nonzero only for the given samples. 
Hence, for the deterministic case of having a finite set of observation samples, we replace the expectation in~\fref{eq:gint} by a sum and state the sparsity measure with $\gdet{\bA,\bY}: U(N;\opC) \times \opC^{N\times M} \to \opR$ as follows:
\begin{align}
\gdet{\bA,\bY} \triangleq \sum_{m=1}^{M}   \vecnorm{\bA\bmy_m}_4^4 = \vecnorm{\bA\bY}_4^4.  \label{eq:gdet}
\end{align}
Then, the corresponding optimization problem is given by
\begin{align}
 \underset{ \bA}{\text{maximize}}\,\, &\gdet{ \bA,\bY}  \,\, \text{subject to } \, \bA\in U(N;\opC), \label{eq:obj2} \tag{O2}
\end{align}
which corresponds to the complex-valued equivalent of the real-valued dictionary learning problem of~\cite{zhai20complete}. Hence, we next provide extensions of the results from~\cite{zhai20complete} to the complex domain.

\subsection{Why the \texorpdfstring{$\ell^4$}{TEXT}-Norm?}

We start by establishing key properties of $\ell^4$-norm over $U(N;\opC)$ to make the relation between the sparsity of a unitary matrix and its $\ell^4$-norm more clear; these properties are extensions of the real-valued $O(N;\opR)$ case from~\cite{zhai20complete}.
The underlying idea behind the properties below is that the sparsest unitary matrices are complex permutation matrices.
Concretely, for any $\bA\in U(N;\opC)$, we have that $N \leq \|\bA\|_0 \leq N^2$, and the lower bound is achieved if and only if $\bA\in\CP(N)$.

The following lemma shows that the global maximizers of the $\ell^4$-norm over the unitary group are complex permutation matrices, which are the most sparse unitary matrices. The proof is provided in \fref{app:l4maxbyCP}.
\begin{lem}[Extrema of the $\ell^4$-Norm Over Unitary Group] \label{lem:l4maxbyCP}
For any unitary matrix $\bA\in U(N;\opC)$, we have that $\|\bA\|_4^4\in[1,N]$. %
The upper bound is achieved, i.e., $\|\bA\|_4^4=N$, if and only if $\bA\in \CP(N)$.
\end{lem}

\begin{figure}
\centering
\includegraphics[width=0.3\linewidth]{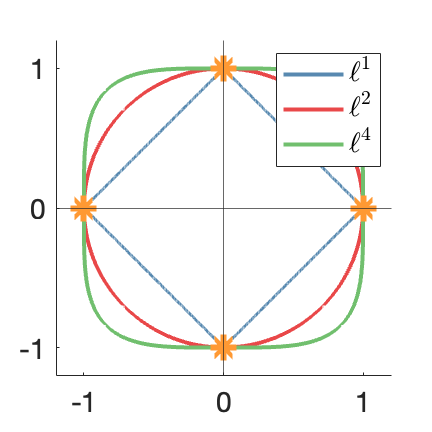}
    \caption{Unit $\ell^1$-, $\ell^2$-, and $\ell^4$-spheres in $\opR^2$.}
    \label{fig:L2sphere}
\end{figure}

\begin{rem}
\fref{fig:L2sphere} provides intuition on why the $\ell^4$-norm over the unitary sphere is a suitable sparsity metric: 
Consider the unitary constraint as being constrained to the unit $\ell^2$-sphere. 
Here, we notice that the $\ell^4$-norm is maximized at the sparsest points on the unit $\ell^2$-sphere designated by the orange stars, where the typical sparsity metric $\ell^1$-norm is minimized. This example illustrates that maximizing $\ell^4$-norm over the unit $\ell^2$-sphere promotes sparsity.
\end{rem}

The following lemma implies that if the $\ell^4$-norm of a matrix is ``close'' to its maximum value, then this matrix is ``close'' to a complex permutation matrix in terms of the Frobenius norm.
The proof is provided in~\fref{app:closetoCP}.

\begin{lem}[Approximate Maxima of $\ell^4$-Norm Over Unitary Group] \label{lem:closetoCP}
Suppose $\bW$ is a unitary matrix, i.e., $\bW \in U(N;\opC)$. If $\frac{1}{N}\vecnorm{\bW}_4^4\geq 1-\epsilon$, $\forall\epsilon\in \!\left[0,\frac{1}{N}\!\left( 1-\frac{1}{\sqrt{2}} \right)\!   \right)$, then there exists $\bP\in \CP(N)$, such that
\begin{align}
   \frac{1}{N}\vecnorm{\bW-\bP}_{\Fro}^2 \leq 2\epsilon. \label{eq:closetoCPbyepsilon}
\end{align}
\end{lem}

We now provide some basic properties of the $\ell^4$-norm over the unitary group $U(N; \opC)$, as extensions of the $O(N;\opR)$ case from~\cite{zhai20complete}.
We define   
$h(\bA) \define g(\bA,\bI_N)$ 
by setting $\bY=\bI_N$ from~\fref{eq:gdet} and rewrite the optimization problem in~\fref{eq:obj2}  as follows:
\begin{align}
\underset{ \bA}{\text{maximize}}\,\, h( \bA)\, \text{ subject to }   \bA\in U(N;\opC). \tag{O3}\label{eq:obj_A}
\end{align}
We utilize $h(\bA)$ in the properties discussed below.

The unitary group is a special case of the complex Stiefel manifold with tangent space given by~\cite{absil2009optimization} 
\begin{equation}
    T_\bW U(N;\opC) \define \{\bZ\in\opC^{N\times N} : \bZ^{\Herm}\bW + \bW^{\Herm}\bZ=\bZero_{N\times N}\}, \label{eq:tangentspace}
\end{equation}
where $\bW\in\opC^{N\times N}$.
The projection operation $\projop{T_\bW U(N;\opC)}:\opC^{N\times N}\to T_\bW U(N;\opC)$ onto the tangent space of $U(N;\opC)$ is defined as \cite{edelman1998geometry} 
\begin{align}
    \proj{T_\bW U(N;\opC)}{\bZ} \define \bZ - \frac{1}{2}\bW(\bW^{\Herm}\bZ+\bZ^{\Herm}\bW) 
    &= \frac{1}{2}(\bZ-\bW\bZ^{\Herm}\bW).
\end{align}
We denote the gradient of $h(\bW)$ with respect to $\bW$  with $\nabla_\bW h(\bW)$, and the Riemannian gradient of $h(\bW)$ with respect to $\bW$ on $T_\bW U(N;\opC)$ with $\grad h(\bW)$.
Hence, we can formulate the Riemannian gradient of $\bW$ on $T_\bW U(N;\opC)$ as below:
\begin{align}
   \grad h(\bW) &= \proj{T_\bW U(N;\opC)}{\nabla_\bW h(\bW)} = \frac{1}{2}(\nabla_\bW h(\bW) - \bW(\nabla_\bW h(\bW))^{\Herm}\bW). \label{eq:riemanniangradientTW}
\end{align}
We utilize  \fref{eq:tangentspace}-\fref{eq:riemanniangradientTW} in order to analyze the critical points of $h(\bW)$ in the following two lemmas.
\begin{lem} 
\label{lem:C1}  
The critical points of $h(\bW)$ on manifold $U(N;\opC)$ satisfy the following condition:
\begin{equation}
    \bW^{\Herm}(|\bW|^{\circ2}\circ\bW) = (|\bW|^{\circ2}\circ\bW)^{\Herm}\bW.
\end{equation}
\end{lem}

The proof of \fref{lem:C1} is provided in \fref{app:C1}.
\fref{lem:C1} implies that for all $ \bW\in U(N;\opC)$, the condition for the critical points of $\ell^4$-norm on $U(N;\opC)$ is given by the following set of equations:
\begin{align}
    \begin{cases}
    \,\,\bW^{\Herm}(|\bW|^{\circ2}\circ\bW) = (|\bW|^{\circ2}\circ\bW)^{\Herm}\bW\\
    \,\,\bW^{\Herm} \bW = \bI.
    \end{cases}
\end{align}
We utilize \fref{lem:C1} to identify the critical points of $h(\bW)$ in \fref{lem:C2} and also in \fref{sec:analysis_msp}.
\begin{lem}  
\label{lem:C2} 
All global maximizers of $\ell^4$-norm over the unitary group are isolated critical points.
\end{lem}
The proof of \fref{lem:C2} is provided in~\fref{app:C2}.
\fref{lem:C2} implies that the global maximizers of the $\ell^4$-norm objective over the unitary group do not form continuous families of solutions; this suggests that small perturbations of a maximizer will not remain maximizers, which is useful for both theoretical guarantees and practical convergence of optimization algorithms.

\section{Learning a Sparsifying Transform}
\label{sec:learning}
We now propose two algorithms to approximately solve \fref{eq:obj1};
extensions to approximately solve~\fref{eq:obj2} are trivial by replacing the expectations by sums as we have done in \fref{eq:gdet}.
We keep our explanations general while introducing our algorithms, and discuss concrete  applications  {for sparsifying channel vectors} in Sections \ref{sec:optimality} and \ref{sec:results}.

\subsection{MSP: Matching, Stretching, and Projection}
We propose an MSP algorithm variant that (i) extends the real-valued method in~\cite[Alg.~2]{zhai20complete} to our complex-valued model and (ii) enables the use of a stochastic model instead of a finite set of observation samples. 
In essence, the MSP algorithm performs a projected gradient ascent (PGA) in the objective~\fref{eq:obj1} with an infinite step size.
In MSP, in each iteration $t$, we \textit{match} the estimate~$\bA_t$ to the observation~$\vecy$, \textit{stretch} all entries of $\bA_t\vecy$ with the cubic function in the gradient $|\bA_t\vecy|^{\circ2}\circ(\bA_t\vecy)$, and \textit{project} it back onto the unitary group, as explained in~\cite{zhai20complete}. 
We utilize the following lemma for the projection onto the unitary group.
\begin{lem}[Projection Onto the Unitary Group]
\label{lem:projU}
For all $\bA\in\opC^{N\times N}$, the unitary matrix which has minimum distance in Frobenius norm to $\bA$ is the following: 
\begin{align}
    \proj{U(N;\opC)}{\bA} \define \underset{\bM\in U(N;\opC)}{\argmin} \vecnorm{\bM-\bA}_{\Fro}^2 = \bU\bV^{\Herm}.
\end{align}
\end{lem}
The proof of \fref{lem:projU} is provided in \fref{app:projU}.
Utilizing the projection above, the resulting MSP procedure to solve the optimization problem in~\fref{eq:obj1} is summarized in~\fref{alg:msp_int}.
\begin{oframed}
\vspace{-0.3cm}
\begin{customalg}{1.1}[MSP]\label{alg:msp_int}
Initialize $\bA_0\in U(N;\opC)$. For every iteration $t=0,1,\dots,$ until convergence, compute the gradient of the objective with respect to $\bA_t$: 
\begin{align}
&\nabla_{\bA_t} \gint{\bA_t } \define \Ex{\bmy} {2\big(|\bA_t\vecy |^{\circ2}\circ(\bA_t\vecy )\big) \vecy^{\Herm} }, \label{eq:msp_int_grad}
\end{align}
and project the gradient onto the unitary group:
\begin{equation}
\bA_{t+1} = \proj{U(N;\opC)}{\nabla_{\bA_t} \gint{\bA_t }}.\! \label{eq:msp_proj}
\end{equation}
\end{customalg}
\vspace{-0.3cm}
\end{oframed}

\subsection{Analysis of the MSP Algorithm}
\label{sec:analysis_msp}
We now provide the convergence analysis of a simplified version of the MSP algorithm over the unitary group in Lemmas~\ref{lem:C3}--\ref{lem:C5} by extending the analysis of~\cite{zhai20complete} for the orthogonal group.
Furthermore, we provide a sufficient condition for the fixed points of~\fref{alg:msp_int} with Lemmata~\ref{lem:UVH} and \ref{lem:fpsofMSP}.

Our convergence analysis  builds on the optimization problem in~\fref{eq:obj_A} with the objective function $h(\bA) =\|\bA\|_4^4 =\gdet{\bA,\bI_N}$.
Thereby, we use a slightly modified version of \fref{alg:msp_int}  by replacing  $\gint{\bA_t }$ and $\nabla_{\bA_t} h(\bA_t) $ by $h(\bA_t)$ and $\nabla_{\bA_t} \gint{\bA_t }  \define {2\big(|\bA_t |^{\circ2}\circ\bA_t\big)}$, respectively.
We write the modified version of the algorithm below in \fref{alg:msp_yless} for the sake of completeness.
\begin{oframed}
\vspace{-0.3cm}
\begin{customalg}{1.2} \label{alg:msp_yless}
Initialize $\bA_0\in U(N;\opC)$. For every iteration $t=0,1,\dots,$ until convergence, compute the gradient of the objective with respect to $\bA_t$: 
\begin{align}
&\nabla_{\bA_t} h(\bA_t) \define {2\big(|\bA_t |^{\circ2}\circ\bA_t\big)}, \label{eq:msp_yless_grad}
\end{align}
and project the gradient onto the unitary group:
\begin{equation}
\bA_{t+1} = \proj{U(N;\opC)}{\nabla_{\bA_t} h(\bA_t)}.\! 
\end{equation}
\end{customalg}
\vspace{-0.3cm}
\end{oframed}
Note that $h(\bA)$ is invariant to unitary rotations and complex phase shifts on $\bA$;
hence, without loss of generality, we wish to show for the optimization problem in~\fref{eq:obj_A} that \fref{alg:msp_yless} converges to some complex permutation matrix in $\CP(N)$ for any randomly initialized $\bA_0$ on $U(N; \opC)$ with probability $1$.

The following lemma establishes properties of the critical points of the $\ell^4$-objective \fref{eq:obj_A}, and \fref{lem:C4} shows that PGA with any fixed step size $\alpha$ (even $\alpha = +\infty$) finds critical points of~\fref{eq:obj_A}.
Then, \fref{lem:C5} shows that \fref{alg:msp_yless} achieves cubic convergence rate locally, which is much faster than gradient-descent methods. 
\begin{lem}[Fixed Point of \fref{alg:msp_yless}] \label{lem:C3}
Given $\bW\in U(N;\opC)$, $\bW$ is a fixed point of \fref{alg:msp_yless} if and only if $\bW$ is a critical point of the $\ell^4$-norm over $U(N;\opC)$.
\end{lem}
The proof of \fref{lem:C3} is provided in \fref{app:C3}.
\fref{lem:C3} implies that \fref{alg:msp_yless} only converges to the critical points of the $\ell^4$-norm over $U(N;\opC)$.
\begin{lem}[Convergence of PGA with Arbitrary Step Size] \label{lem:C4}
An iterative projected gradient ascent (PGA) algorithm with any fixed step size $\alpha>0$ ($\alpha$ can be $+\infty$, in which case PGA is equivalent to \fref{alg:msp_yless}) finds a saddle point of the optimization problem: $\underset{\bA\in U(N;\opC)}{\max} \vecnorm{\bA}_4^4$.
\end{lem}
The proof of \fref{lem:C4} is provided in~\fref{app:C4}.
\fref{lem:C4} implies that PGA with any step size $\alpha>0$ (including $\alpha=+\infty$) finds a critical point of $\vecnorm{\bA}_4^4$ over $U(N;\opC)$.

\begin{lem}[Cubic Convergence Rate around Global Maximizers] \label{lem:C5}
Given a unitary matrix $\bA\in U(N;\opC)$, let $\bA'$ denote the output of \fref{alg:msp_yless} after one iteration:
$\bA' = \bU\bV^{\Herm}$, where $\bU\bSigma\bV^{\Herm} = \SVD{\bA^{\circ2}\circ \bA}$.
If $\vecnorm{\bA-\bC}_{\Fro}^2 = \epsilon$, for $\epsilon <  0.394$, then we have that $\vecnorm{\bA'-\bC'}_{\Fro}^2 < \vecnorm{\bA-\bC}_{\Fro}^2$ and $\vecnorm{\bA'-\bC'}_{\Fro}^2 < O(\epsilon^3)$, where $\bC,\bC'\in \CP(N)$ are diagonal complex permutation matrices such that $c_{i,i}=e^{j\angle A_{i,i}}$ and $c_{i,i}'=e^{j\angle A_{i,i}'}$.
\end{lem}
The proof of \fref{lem:C5} is provided in~\fref{app:C5}.
\fref{lem:C5} shows that if the initial unitary matrix $\bA$ is close enough to a complex permutation matrix, then \fref{alg:msp_yless} achieves convergence to a complex permutation at a cubic rate.
Since the algorithm is permutation invariant, we assumed---without loss of generality---that the target complex permutation is a diagonal complex permutation matrix $\bC\in\opC^{N\times N}$.

So far, we have provided a convergence analysis of~\fref{alg:msp_yless}.
We next provide a lemma that helps identify the fixed points of~\fref{alg:msp_int}.

\begin{lem} 
\label{lem:UVH}
Let $\bA\in U(N;\opC)$ be a unitary matrix with $\SVD{\bA}=\bU\bSigma\bV^{\Herm}$ and $\bD\in\opR^{N\times N}$ be a diagonal matrix.
Suppose we have the matrices $\bA_1 = \bD\bA$ and $\bA_2 = \bA\bD$ with $\SVD{\bA_1}=\bU_1\bSigma_1\bV_1^{\Herm}$ and $\SVD{\bA_2}=\bU_2\bSigma_2\bV_2^{\Herm}$. Then, 
there exist SVDs for $\bA_1$ and $\bA_2$ such that $\bSigma_1 = \bSigma_2 = \bD$ and $\bA = \bU\bV^{\Herm} = \bU_1\bV_1^{\Herm} = \bU_2\bV_2^{\Herm}$.
\end{lem}
The proof of \fref{lem:UVH} is provided in~\fref{app:UVH}.
We now use~\fref{lem:UVH} to provide a sufficient condition for the fixed points of \fref{alg:msp_int}.
\begin{lem}[Fixed points of MSP]
\label{lem:fpsofMSP}
A unitary matrix $\bA\in U(N;\opC)$ is a fixed point of \fref{alg:msp_int} if there exists a diagonal matrix $\bD\in\opR^{N\times N}$ such that $\nabla_\bA \gint{\bA } = \bD\bA$ or $\nabla_\bA \gint{\bA } = \bA\bD$.
\end{lem}
The proof of~\fref{lem:fpsofMSP} immediately follows from~\fref{lem:projU} and \fref{lem:UVH}.
In \fref{sec:opt_dft}, we will use~\fref{lem:fpsofMSP} to examine the DFT's optimality for sparsifying channel vectors. 

\begin{rem}
Establishing local optimality of a fixed point of the MSP algorithm is challenging: The optimization problem in~\fref{eq:obj1} is over the unitary group, which is a Riemannian manifold; this requires one to show that the Riemannian gradient vanishes and the Riemannian Hessian is negative semidefinite at a certain point to prove that this point is a local maximum~\cite{absil2009optimization}. 
While such an analysis was feasible for the problem setup in~\cite{liu24ofdm}, the same approach is not applicable here. 
We can get one step closer to studying local optimality via the procedure detailed next.
\end{rem}

\subsection{CA: Coordinate Ascent}
\label{sec:ca}

We now propose an alternative approach to approximately solve the optimization problems in~\fref{eq:obj1} and \fref{eq:obj2}. 
Here, we make use of the fact that every unitary matrix can be written as a sequence of real-valued Givens rotations and complex phase shifts, as shown in~\cite{christoph_givens,hassibi2002cayley}. 
Let $\bG(i,k,\alpha_{i,k})\in\opC^{N\times N}$ for $i>k$ denote the real-valued Givens rotation matrix of the form $G_{i,i}=G_{k,k}=\cos(\alpha_{i,k}),\,G_{i,k}= - G_{k,i}=\sin(\alpha_{i,k}),\,G_{\ell,\ell}=1,\ell\neq i$, and $G_{\ell,m}=0$ otherwise.
Multiplying a matrix with $\bG(i,k,\alpha_{i,k})$ from the left amounts to a counterclockwise rotation of $\alpha_{i,k}$ radians in the $(i,k)$ coordinate plane.
Let $\bR(k,\beta_k)\in\opC^{N\times N}$ denote a diagonal, complex-valued phase shift matrix of the form $R_{k,k} = e^{j\beta_k}$ and $R_{\ell,\ell} = 1$ if $\ell\neq k$.
We rewrite the optimization variable $\bA$ in~\fref{eq:obj1}  as $\bA= \prod_{k=N}^{1}  \prod_{i=k+1}^{N} \bG(i,k,\alpha_{i,k}) \prod_{\ell=k}^{N} \bR(\ell,\beta_{i,\ell})$.
This decomposition results in the following unconstrained optimization problem that is equivalent to~\fref{eq:obj1}:
\begin{align}
\underset{ \substack{ \{\alpha_{i,k}\in \opR: i,k\in\setint{N},i>k\}, \\ \{\beta_{k,\ell} \in \opR: k,\ell\in\setint{N},\ell\geq k \} }}
{\text{maximize}}\,\, & \Ex{\bmy}{ \left\| \left(   \prod_{k=N}^{1}  \prod_{i=k+1}^{N} \bG(i,k,\alpha_{i,k}) \prod_{\ell=k}^{N} \bR(\ell,\beta_{k,\ell})\right)\bmy  \right\|_4^4  }\!.  \tag{O4} \label{eq:ca_obj1}
\end{align} 
While the optimization problem in~\fref{eq:ca_obj1} is unconstrained as opposed to~\fref{eq:obj1}, it is still difficult to directly solve as the objective function is nonconvex with respect to the optimization variables.
To this end, we propose a coordinate ascent (CA)-based algorithm, \fref{alg:ca_int}, to iteratively find an approximate solution to~\fref{eq:ca_obj1};
this approach avoids a projection step as each iteration update preserves unitarity.
\begin{oframed}
\vspace{-0.3cm}
\begin{customalg}{2}[CA]\label{alg:ca_int}
Initialize $\bA_0 \in U(N;\opC)$. 
For every iteration $t=0,1,\dots,$ until convergence, select an index pair $(i,k)$ such that  $i,k\in\setint{N}, i>k$, and find 
\begin{align}
\alpha_{i,k} & \in \underset{\tilde\alpha}{\argmax}\,\, \Ex{\bmy}{ \vecnorm{\bG(i,k,\tilde\alpha)\bA_t\vecy }_4^4 }, \label{eq:ca_alpha} \\
\{\beta_i,\beta_k\} & \in \underset{\tilde\beta_i,\tilde\beta_k }{\argmax} \,\,\Ex{\bmy}{\vecnorm{\bG(i,k,\alpha_{i,k})\bR(i,\tilde\beta_i)\bR(k,\tilde\beta_k)\bA_t\vecy }_4^4 }, \label{eq:ca_beta}
\end{align}	
and apply the update
\begin{align}
\bA_{t+1} &=  \bG(i,k,\alpha_{i,k})\bR(i,\beta_i)\bR(k,\beta_k)\bA_t. \label{eq:ca_update}
\end{align}
\end{customalg}
\vspace{-0.3cm}
\end{oframed}
The idea behind this CA algorithm is to sequentially update a pair of indices of a given initial unitary matrix by multiplying this matrix from the right by Givens rotation matrices and phase {shift} matrices to improve the $\ell^4$-norm objective.
Note that since the $\ell^4$-norm is invariant to complex phase shifts, in each iteration, we first optimize for $\alpha_{i,k}$, and then for $\beta_i$ and $\beta_k$ accordingly. 
Since the multiplication of Givens rotation matrices and phase shift matrices with a unitary matrix is still unitary, the algorithm preserves unitarity in each iteration and avoids a projection onto the unitary group altogether.
This key property enables us to analyze local optimality at each CA iteration by directly analyzing the first and second derivatives.

\begin{rem}
The complexity of the CA algorithm is typically higher than that of MSP, as we have to iterate through all $(i,k)$ index pairs \textit{at least once} and solve the optimization problems in~\fref{eq:ca_alpha} and \fref{eq:ca_beta} by another iterative algorithm, such as gradient ascent or line search, for each $(i,k)$. 
The advantage of the CA algorithm is that it enables one to establish local optimality, since at each iteration, we solve an unconstrained optimization problem with scalar optimization variables (i.e., Givens rotation angles). In contrast, the MSP algorithm solves a constrained problem in which the optimization variable is an $N\times N$ matrix,
which makes an optimality analysis difficult.
Optimality criteria at each iteration of the CA algorithm are detailed next.
\end{rem}

\subsection{Analysis of CA}

It is trivial that \fref{alg:ca_int} is monotonically nondecreasing in its objective in every iteration. 
While we cannot make global convergence claims, as the objective function is nonconvex, we can study the fixed points of the algorithm along with the local extrema of the objective function in~\fref{eq:ca_alpha} at each iteration;
these properties are analyzed by the following lemma.
\begin{lem} \label{lem:ca_opt} 
Let the matrix~$\bG$ denote the Givens rotation matrix as defined in~\fref{sec:ca} and $\bmx\define\bA\bmy$ for a unitary matrix $\bA\in U(N;\opC)$.
The  matrix $\bA$  is a fixed point of~\fref{alg:ca_int} 
and the function $\vecnorm{\bG(i,k,\alpha_{i,k})\bA\vecy}_4^4$ from~\fref{eq:ca_alpha} attains a local maximum at $\alpha_{i,k}=0$ if and only if the two following conditions hold for all $ i,k\in \setint{N} ,\, i>k$:
\begin{align}    
\textnormal{(i)} \quad 
&\Ex{\bmy}{\pder{}{\alpha} \big(\vecnorm{\bG(i,k,\alpha)\bA\vecy}_4^4\big)}\bigg|_{\alpha=0}
= 4\Ex{\bmx}{\Re\{ x_ix_k^\star\} (|x_i|^2 - |x_k|^2)} = 0 ,\label{eq:first_deriv}  \\
\textnormal{(ii)}  \quad 
&\Ex{\bmy}{\pder{^2}{\alpha^2} \big(\vecnorm{\bG(i,k,\alpha)\bA\vecy}_4^4\big)}\bigg|_{\alpha=0} 
= 4\Ex{\bmx}{  2\Re\{x_k^2(x_i^*)^2\} + 4|x_i|^2|x_k|^2 - |x_k|^4 - |x_i|^4  }< 0. \label{eq:second_deriv} 
\end{align}
\end{lem}
The proof of~\fref{lem:ca_opt} immediately follows from the first- and second-derivative tests~\cite{strang1991calculus}, the equality that $\bA=\bG(i,k,0)\bA$, and simplifying the derivative expressions for $\alpha=0$. 
We will utilize this lemma in Sections~\ref{sec:opt_dft} and~\ref{sec:opt_dct} when examining DFT's optimality for sparsifying channel vectors and DCT's optimality for sparsifying real-valued sinusoids, respectively. 
 

\section{Optimality of Known Transforms}
\label{sec:optimality}

We now discuss concrete applications of our algorithms in~\fref{sec:learning} to examine  optimality of two common transforms: the (i) DFT and (ii) DCT. 

\subsection{Discrete Fourier Transform}
\label{sec:opt_dft}

Since our prime goal is to sparsify channel vectors of mmWave systems as mentioned in~\fref{sec:prereqs},  
we adopt a stochastic data model $\bmy\in\opC^B$ according to~\fref{eq:h_sum} as follows:
\begin{equation} \label{eq:ejw}
    y_b =\sum_{\ell=1}^L c_\ell e^{j\Omega_\ell (b-1)}.
\end{equation}  
Here, $\Omega_\ell\sim \Unif(0,2\pi)$ and i.i.d. for $\ell=1,\dots,L$.
This data model corresponds to a multipath channel model with a uniform distribution over the angular frequency associated with each path.\footnote{Another stochastic data model could use $e^{j\pi\sin{\Phi_\ell}\vecb}$,  $\Phi_\ell\sim \Unif(0,2\pi)$, which assumes a uniform distribution over the incidence angle~\cite{tse} associated with each path. However, we use the model in \fref{eq:ejw} to facilitate our theoretical analysis.}  
Based on the vast literature on beamspace processing (see, e.g.,~\cite{Sayeed13GLOBECOM, schniter14, gao16bs, chen17bscs, ma18,abdelghany2019precoding,mirfarshbafan19a,seyedicassp20,mahdavi20beamspace,mirfarshbafan21tcas,gonultacs2021hardware,osinsky21,dai22jsips,he23twc,alimo24}), it is known that the DFT is a good candidate to sparsify channel vectors. Thus, we use the DFT to initialize both our MSP and CA algorithms to (i) examine the DFT's optimality and (ii) find potentially better sparsifying transforms. 
An optimality analysis of DFT for both algorithms is detailed next.

\subsubsection{Analysis with MSP}
\label{sec:opt_dft_msp}
We establish that the DFT matrix is a fixed point of the MSP algorithm for the stochastic data model in~\fref{eq:ejw} with the following result.
\begin{theorem} 
\label{thm:dft_msp_fp}
For the stochastic data model $\vecy $ as given in~\fref{eq:ejw},
the DFT matrix $\bF_B$ is a fixed point of~\fref{alg:msp_int}.
\end{theorem}
The proof of~\fref{thm:dft_msp_fp} is provided in~\fref{app:dft_msp_fp}.
\fref{thm:dft_msp_fp} implies that \fref{alg:msp_int} is unable to find any unitary transform that is better at sparsifying the multipath channel model from~\fref{eq:ejw}.
Note that this analysis does not establish whether the DFT matrix corresponds to a saddle point or a local maximum of the objective function in~\fref{eq:obj1}. 
Nonetheless, we can use the proposed  CA algorithm to analyze local maxima.

\subsubsection{Analysis with CA}
\label{sec:opt_dft_ca}

For the stochastic data model from~\fref{eq:ejw} with a free-space line-of-sight channel model (i.e., $L=1$), the following theorem establishes that the DFT matrix is a fixed point of~\fref{alg:ca_int} and CA achieves a local maximum of the objective function in~\fref{eq:ca_alpha}. 
\begin{theorem} 
\label{thm:dft_opt}
For the stochastic data model $\vecy $ as given in~\fref{eq:ejw} with $L=1$,
the DFT matrix $\bF_B$ is a fixed point of~\fref{alg:ca_int} and the objective function
$\vecnorm{\bG(i,k,\alpha_{i,k})\bA\vecy}_4^4$ from~\fref{eq:ca_alpha} attains a local maximum for all $ i,k\in \setint{B} ,\, i>k$.
\end{theorem}
The proof of~\fref{thm:dft_opt} is provided in~\fref{app:dft_opt}.
Although we cannot formally prove whether 
the function $\vecnorm{\bG(i,k,\alpha_{i,k})\bF_B\vecy}_4^4$ from~\fref{eq:ca_alpha} attains a global maximum at $\alpha_{i,k}=0$
we have observed no solution, upon perturbations and random initialization, that achieves a higher value of the objective function in~\fref{eq:gint} for the stochastic data model $\vecy $ in~\fref{eq:ejw} with $L=1$.
We leave establishing global optimality to future work.  
    
\subsection{Discrete Cosine Transform}
\label{sec:opt_dct}

The above results on the DFT's optimality to sparsify complex sinusoids may seem obvious. Analogously, one would expect that the DCT---as the DFT's real-valued counterpart---to be similarly optimal for the following stochastic data model $\vecy \in\opR^{B}$ in the form of a real-valued sinusoid such that
\begin{equation} \label{eq:cosw}
    y_b =\cos{(\Omega (b-1) +\Phi)}, \,\, b=1,\ldots,B,
\end{equation}
where $\Omega\sim \Unif(0,2\pi)$ and $\Phi\sim \Unif(0,2\pi)$.
However, as we detail below, the DCT is surprisingly suboptimal for various instances of $B$---this affirms that optimality of the DFT for a complex-valued sinusoid-based data model with any vector length $B$ is not obvious.
If the DCT matrix is not a fixed point of either one of the MSP or CA algorithms, then the algorithm can find a sparsifying transform that increases the objective function, which implies that the DCT is suboptimal in the $\ell^4$-norm sense.

\subsubsection{Analysis with CA}
\label{sec:dct_opt}

Let us denote the $B\times B$ DCT-II matrix with $\bC_B$ in this section.\footnote{Although we only provide the simplified analytical expression for DCT-II in~\fref{app:dct_subopt}, we numerically verified that our result is valid for the DCT types I, III, and IV as well.}
We simplify the first derivative from~\fref{eq:first_deriv} with $\bA$ set to $\bC_B$ and the stochastic data model~$\bmy$ as given in~\fref{eq:cosw}; the resulting closed-form expression is provided in~\fref{app:dct_subopt}. 
While showing that this closed-form expression is nonzero for \textit{any} $B>2$ is difficult, 
our numerical evaluations of this expression for $B=3,\dots,200$ revealed that there exists $(i,k)$ such that $ \pder{\int_0^{2\pi}\vecnorm{\bG(i,k,\alpha)\vecx(\omega)}_4^4\td\omega}{\alpha}\bigg|_{\alpha=0} \neq 0$, meaning that $\bC_B$ is not a fixed point of CA and not an optimal sparsifier for the data model defined in~\fref{eq:cosw}.  While we have not numerically calculated the expression  for larger values of $B$, we expect the same result.

\begin{rem}
Given that the DFT is locally optimal for complex-valued sinusoids, one would expect the DCT to be optimal for real-valued sinusoids---so this result has been surprising. 
However, we note that the DCT is \textit{close to} optimal because the new transforms found by the MSP or CA algorithms have only a minor improvement in the objective function of~\fref{eq:gint} compared to the DCT. 
\end{rem}


\graphicspath{ {./figures/} }
\newcommand{\figsize}{0.42}
 
\section{Numerical Results}
\label{sec:results}

We now utilize~\fref{alg:msp_int} to learn a sparsifying transform from a given, finite-size data set of channel vectors for two experiments: (i) synthetically-generated and (ii) real-world (measured) channel vectors. 
We note that in these cases,  our theoretical results on the optimality of the DFT are no longer guaranteed since the channel vector distribution in the datasets do not follow the data model from~\fref{eq:ejw}, which has uniform distribution over the angular frequency associated with each path. 
To be specific, in (i), the angular frequencies associated with each path are not uniformly distributed as this is a finite dataset, and in (ii), even the channel model from~\fref{eq:h_sum} does not hold due to hardware impairments.

In all of our experiments, we split the channel matrices into training and test sets; 
we use training set to learn a sparsifying transform; then, on the test set, we measure the $\ell^4$-norm and simulate the uncoded bit error rate (BER) with respect to signal-to-noise ratio (SNR) for various sparsity-exploiting algorithms.
Specifically, we use (i) the beamspace channel estimation (BEACHES) algorithm from~\cite{mirfarshbafan19a} and (ii) the beamspace largest-entry (LE) data detector  from~\cite{seyedicassp20} with density coefficient $0.125$, which processes only $12.5$\% of the entries of the sparse data to reduce complexity. 
As a baseline, we also include an antenna-domain linear least-squares minimum mean-square error (LMMSE) data detector, which does not rely on sparsity, while still using BEACHES to denoise the channel vectors. 

\begin{figure}[tp]
\centering
\includegraphics[width=.5\columnwidth]{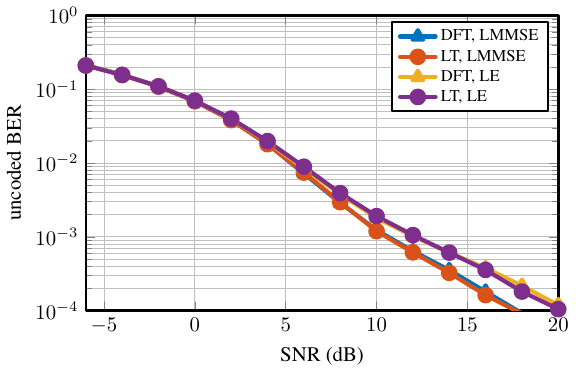}
\vspace{-0.4cm}
\caption{Uncoded BER for beamspace processing with the DFT or learned transform (LT) using synthetic QuaDRiGa channels with LMMSE and  the sparsity-exploiting LE detector, $B=256$ BS antennas, and $U=16$ UEs.}
\label{fig:ber_quad}
\vspace{-0.1cm}
\end{figure}
\begin{figure}
\centering
\includegraphics[width=.5\columnwidth]{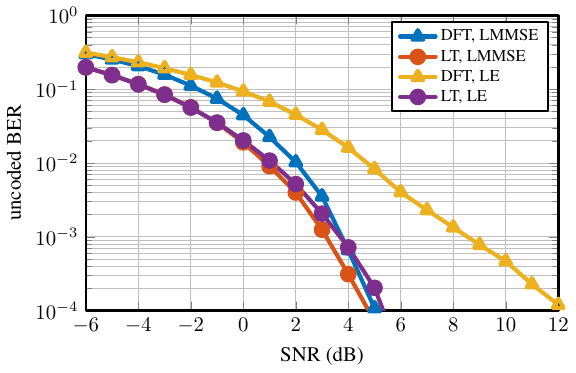}
\vspace{-0.4cm}
\caption{Uncoded BER  for beamspace processing with the DFT or learned transform (LT) using measured channel vectors~\cite{gauger2020massive} with LMMSE and the LE detector, $B=64$ BS antennas (with eight malfunctioning), and $U=1$ UE.}
\label{fig:ber_ctw}
\vspace{-0.1cm}
\end{figure}

\subsection{Synthetic Channel Vectors}
\label{sec:results_synthetic}
We first simulate line-of-sight (LoS) channel conditions using the QuaDRiGa mmMAGIC UMi model\cite{QuaDRiGa} at a carrier frequency of 60\,GHz with a ULA with $\lambda/2$ antenna spacing.
{While representing an idealistic scenario, the channel vectors in this dataset do not directly follow the theoretical model in~\fref{eq:ejw} since the dataset has a finite number of channel vectors and the distribution of the angular frequencies $\Omega_\ell$ associated with each path in a channel vector is no longer uniform as $\Unif(0,2\pi)$.}
We generate channel matrices for a mmWave massive MIMO system with $B=256$ BS antennas and $U = 16$ single-antenna UEs.
The UEs are placed randomly in a $120^\circ$ circular sector around the BS between a distance of 10\,m and 110\,m, and we assume a minimum UE separation of  $5^\circ$.
We add BS-side power control so that the UE with the highest received power has at most $6$\,dB more than the weakest UE.
\fref{fig:ber_quad} shows BER results for this channel vector set using the DFT and the learned transform (LT).
We observe that the LT has only a marginal BER advantage compared to the DFT. Moreover, the $\ell^4$-norm of the test set in beamspace domain with LT is only $18\%$ higher than that of the DFT, which can be interpreted as only a $2\%$ difference in magnitude of the peaks in the sparse domain.
This result demonstrates that the DFT is (i) no longer optimal but (ii) remains an excellent sparsifier for simulated mmWave LoS channels, which is in congruence with our proof of optimality of the DFT for the approximate model used in~\fref{eq:ejw}. 
Consequently, we feel that it is not worth learning a new sparsifying transform for minor improvements given that the DFT can be applied with a fast Fourier transform, whereas the learned transform does not necessarily have a fast algorithm. 

We have also experimented with ``semi-uniform arrays,'' motivated by the challenges encountered in designing of large antenna arrays for terahertz (THz) communication applications; see, e.g.,~\cite{farid135ghz}.
Here, the usual $\lambda/2$ inter-element spacing cannot be preserved between tiles of subarrays, leading to arrays consisting of uniform subarrays where the spacing between these subarrays is larger than the element spacing within each subarray. For example, the inter-subarray spacing can be~$\lambda$ while the inter-element spacing within each subarray is $\lambda/2$.  
For such an array, we created channel vectors in a similar fashion to above and learned a sparsifying transform. 
Our results revealed that the DFT's performance with this semi-uniform array was close to the uniform array case. Therefore, the learned transform could provide only slightly better error performance than the DFT and---once again---it does not seem to be worth learning new transforms.

\subsection{Real-World Measured Channel Vectors}
We now show results for \emph{measured} channel vectors provided for the IEEE Communications Theory Workshop Localization Competition~\cite{gauger2020massive}.
The channel measurements in this dataset significantly deviate from the theoretical model in~\fref{eq:ejw} due to having a rectangular array with non-ideal hardware instead of an ideal ULA.
These channel measurements are based on single-UE transmission to a BS with an $8\times 8$ square antenna array with $\lambda/2$ spacing at a carrier frequency of 1.27\,GHz. 
Eight BS antennas were malfunctioning and their outputs were excluded from the dataset; we applied zero-padding for such missing entries. 
For beamspace processing with rectangular arrays, one would typically apply a two-dimensional DFT on this data as follows:  
Vectorize the data to have vectors of size $64$, then multiply with $\bF_8\otimes\bF_8$.
Instead, we use these  
vectors of length 64 as a training set to learn a sparsifying transform.

\fref{fig:ber_ctw} shows BER results for this channel vector set using the DFT and the learned transform (LT). 
We observe that the LT can achieve a target BER of $0.1$\% at approximately 5\,dB and 1\,dB smaller SNR than the DFT with the LE and LMMSE detectors, respectively.
Moreover, the $\ell^4$-norm of the test set after applying the LT is up to $4\times$ higher than that of the two-dimensional DFT.
This result demonstrates that, although the (two-dimensional) DFT is well-suited for beamspace processing under idealistic LoS channel conditions, learning new sparsifying transforms enables significant improvements for real-world channels and communication systems suffering from hardware impairments, such as malfunctioning antennas.


\section{Conclusions}
\label{sec:conclusions}

We have formulated an $\ell^4$-norm maximization problem to learn unitary sparsifying transforms by generalizing the real-valued dictionary learning problem in~\cite{zhai20complete} to the complex case. 
We have considered learning from both a stochastic model and a dataset. 
We have proposed two algorithms for this optimization problem: (i) a projected gradient ascent-based algorithm adapted from~\cite[Alg.~2]{zhai20complete} and a novel coordinate ascent algorithm that avoids projection onto the unitary group. 
We have used these algorithms to study the optimality of the DFT for sparsifying channel vectors. 
Our theoretical analysis has revealed that for a multipath channel model, the DFT matrix is a fixed point of the first algorithm. Moreover, for a free-space line-of-sight channel model, the DFT matrix is a fixed point of the second algorithm, where a local maximum of the objective function at each iteration is achieved.   
In order to reveal the nontrivialness of this result, we have proved that the DCT is surprisingly suboptimal for a (real-valued) sinusoid data model.
To demonstrate that the DFT indeed performs well for idealistic mmWave channel models, we have provided simulation results. Our results have shown that learned transforms can significantly improve the BER for real-world measurements with non-ideal hardware. 

Although our focus was on mmWave communication systems, our algorithms are applicable to more general problems on learning sparsifying transforms, potentially including other stochastic data models or optimality claims, which leads to many avenues for future work. It would also be interesting to learn transforms that have a fast algorithm with lower complexity than $O(N^2)$, similar to the fast Fourier transform (FFT).
 

\appendices

\section{Proofs of~\fref{sec:l4setup}}
\label{app:app_prereqs} 

\subsection{Proof of \fref{lem:l4maxbyCP}}
\label{app:l4maxbyCP}
We write an upper bound for $\|\bA\|_4^4= \sum_{k=1}^N\sum_{\ell=1}^N |A_{k,\ell}|^4 $ as follows: 
\begin{align}
    &\|\bA\|_4^4  \leq \sum_{k=1}^N\sum_{\ell=1}^N |A_{k,\ell}|^4 + 2\sum_{k=1}^N \!\sum_{1\leq \ell_1 < \ell_2\leq N} |A_{k,\ell_1}|^2|A_{k,\ell_2}|^2 
    =\sum_{k=1}^N \!\left(\sum_{\ell=1}^N |A_{k,\ell}|^2\right)^2 = N.
\end{align}  
The upper bound is reached if and only if
    $A_{k,\ell_1}A_{k,\ell_2} = 0,  \forall k,\ell_1\neq \ell_2 \in \setint{N}$,
i.e., when each row of $\bA$ has only one non-zero entry.
Combined with the condition that $\bA$ is unitary, this  implies  that $\bA\in \CP(N)$.

We write a lower bound for $\|\bA\|_4^4$ as follows with $\tilde\bma\define\mathrm{vec}(\bA)$:
\begin{align}
    &\|\bA\|_4^4 =\|\tilde\bma\|_4^4 =  \sum_{k=1}^{N^2}  |\tilde a_{k}|^4 \geq  \frac{1}{N^2} \!\left( \sum_{k=1}^{N^2}  |\tilde a_{k}|^2 \right)^2 = 1.
\end{align}  
By the Cauchy--Schwarz inequality \cite{steele2004cauchy}, the lower bound is reached if and only if~$|\tilde \bma|^{\circ 2}$ and~$\bOne_{N^2}$ are linearly dependent.
Combined with the condition that $\bA$ is unitary, this implies  that $|A_{i,k}|=\frac{1}{\sqrt{N}},\forall i,k$.

\subsection{Proof of \fref{lem:closetoCP}}
\label{app:closetoCP} 
We can rewrite the condition that $\frac{1}{N}\vecnorm{\bW}_4^4\geq 1-\epsilon$ as 
\begin{align}
    \frac{1}{N}\sum_{k=1}^{N} \vecnorm{\vecw_k}_4^4\geq 1-\epsilon,  
\end{align}
where each column vector $\vecw_k$ satisfies $\vecw_k\in\opS^{N-1}$. 
By \fref{lem:multiple_vecs_conv}, we know there exists $ m_1,m_2,\dots,m_N \in
\setint{N}$, such that the following holds for any $\epsilon\in[0,1]$:
\begin{align}
\frac{1}{N}\sum_{k=1}^{N} \vecnorm{ \vecw_k-\vece_{m_k}^{\theta_k}}_2^2 \leq 2\epsilon. \label{eq:twonorm_wminuse}
\end{align}
Here,  $\vece_{m_k}^{\theta_k}\in\opC^N$ with $[\vece_{m_k}^{\theta_k}]_{m_k}=e^{j\theta_k},\theta_k\in[0,2\pi)$, and $[\vece_{m_k}^{\theta_k}]_{\ell}=0,\forall \ell\neq m_k,\ell\in\setint{N}$. 
In fact, from \fref{lem:single_vec_conv}, we know that \fref{eq:twonorm_wminuse} is satisfied by $m_k \in \argmax_{i\in[N]} |W_{i,k}|$ and $\theta_k = \angle W_{m_k,k}$  specifically.
Let $\bP\in\opC^{N\times N}$ such that $\bmp_k = \vece_{m_k}^{\theta_k}$.
So far, we have only deduced that there exists~$\bP$ whose columns have only one nonzero entry with unit magnitude.
We have not yet analyzed if $\bP$ is a complex permutation matrix, which requires that $m_k\neq m_j, \forall k\neq j$, i.e., the sequence $m_1,\dots,m_N$ should have no repetitions;
this is exactly what we prove next.

By simplifying the sum $ \sum_{k=1}^{N}\vecnorm{ \vecw_k-\vece_{m_k}^{\theta_k}}_2^2 = \sum_{k=1}^{N} (2 -2 |W_{m_k,k}|)$, we rewrite~\fref{eq:twonorm_wminuse} as follows:
\begin{align}
\sum_{k=1}^{N} (1 -|W_{m_k,k}|) \leq N\epsilon.\label{eq:twonorm_wminuse_v2}
\end{align}
Let $\setK_i\define \{k\in\setint{N}:m_k = i\}$ be the set of columns of $\bP$ whose nonzero entry is at row $i$ with $c_i\define |\setK_i|$.
Let $\setK_i' \define \{k\in\setK_i: |W_{i,k}| \leq \frac{1}{\sqrt{2}}\}$.
Now, note that if $c_i \geq 2$, then $|\setK_i'|\geq c_i-1$ must hold, since $\sum_{k \in \setK_i} |W_{i,k}|^2 \leq 1 $ by the unitarity of $\bW $.
Let $N_r\define \sum_{i\in\{m_k\}_{k=1}^N} (c_i-1) $ be the total number of repetitions in $m_1,\dots,m_N$ with $0\leq N_r \leq N-1$.
Then, we have that
\begin{align}
     N_r\bigg(1- \frac{1}{\sqrt{2}}\bigg) \leq \sum_{k\in\setK_{m_k}'} (1 -|W_{m_k,k}|)
     \leq \sum_{k=1}^{N} (1 -|W_{m_k,k}|)  \leq N\epsilon < \bigg(1- \frac{1}{\sqrt{2}}\bigg), \label{eq:r_ub1}
\end{align}
From \fref{eq:r_ub1}, we obtain that $N_r<1$;
since $N_r$ is a nonnegative integer, this implies that $N_r=0$, which concludes our proof.

\subsection{Proof of \fref{lem:C1}}
\label{app:C1}

Since $h(\bW)=\vecnorm{\bW}_4^4$, we have that $\nabla_\bW h(\bW) = 2|\bW|^{\circ2}\circ\bW$. Inserting this gradient into~\fref{eq:riemanniangradientTW} yields the following:
\begin{align} 
    &\grad h(\bW) = \frac{1}{2}(2|\bW|^{\circ2}\circ\bW - \bW(2|\bW|^{\circ2}\circ\bW)^{\Herm}\bW)=0 \\
    &\implies |\bW|^{\circ2}\circ\bW = \bW(|\bW|^{\circ2}\circ\bW)^{\Herm}\bW. 
\end{align}
Multiplying the left hand side by $\bW^{\Herm}$ results in  $\bW^{\Herm}(|\bW|^{\circ2}\circ\bW) = (|\bW|^{\circ2}\circ\bW)^{\Herm}\bW$.

\subsection{Proof of \fref{lem:C2}}
\label{app:C2}

We prove that a \textit{diagonal} complex permutation matrix is an isolated critical point without loss of generality, since our objective function is invariant under the complex permutation group.
Let $\bW_0$ be a diagonal matrix such that $[\bW_{0}]_{k,k} = e^{j\theta_k}$ for $k\in\setint{N}$ and  $\theta_k\in[0,2\pi)$, 
and let the critical point conditions, i.e. $\bW_0^{\Herm}(|\bW_0|^{\circ2}\circ\bW_0) = (|\bW_0|^{\circ2}\circ\bW_0)^{\Herm}\bW_0$ and $\bW_0^{\Herm}\bW_0=\bI_N$, hold.
Now let us examine if in every neighborhood of $\bW_0$, there exists some $\bW$ that is a critical point, i.e., if there exists a path $\bW(\cdot): (-\epsilon,\epsilon)\to U(N;\opC)$ around $\bW_0$ such that $\bW(0) = \bW_0$ and the following condition holds:
\begin{align} 
   &\bW(t)^{\Herm}(|\bW(t)|^{\circ2}\circ\bW(t)) = (|\bW(t)|^{\circ2}\circ\bW(t))^{\Herm}\bW(t). \label{eq:constraint1} 
\end{align}
Expanding $\bW(t)$ around $t=0$, we have that 
\begin{equation}
    \bW(t) = \bW_0 + t\bW_1 + t^2\bW_2 + \cdots.
\end{equation}
By inserting $\bW(t)$ into the constraint in \fref{eq:constraint1}, we have that
\begin{align}
&(\bW_0+t\bW_1)^{\Herm} \left((\bW_0+t\bW_1)^{\circ2}\circ(\bW_0+t\bW_1)^*\right)
=\left((\bW_0+t\bW_1)^{\circ2}\circ(\bW_0+t\bW_1)^*\right)^{\Herm} (\bW_0+t\bW_1). \label{eq:constraint1implies}
\end{align}
Simplifying the right-hand side of \fref{eq:constraint1implies} gives
\begin{align}  
&(\bW_0+t\bW_1)^{\Herm} \left((\bW_0+t\bW_1)^{\circ2}\circ(\bW_0+t\bW_1)^*\right)  
= \bI + 2t\bW_0^{\Herm}\circ \bW_1 + t\bW_0 \circ \bW_1^* + t\bW_1^{\Herm}\bW_0,
\end{align} %
Similarly, simplifying the left-hand side of \fref{eq:constraint1implies} gives
\begin{align}   
&\left((\bW_0+t\bW_1)^{\circ2} \circ (\bW_0+t\bW_1)^*\right)^{\Herm} (\bW_0+t\bW_1) 
= \bI + t\bW_0^{\Herm}\bW_1 + 2t \bW_1^{\Herm} \circ \bW_0  + t \bW_0^{\Herm} \circ\bW_1 .
\end{align}
Hence, we simplify \fref{eq:constraint1implies} to the following:
\begin{align}  %
&   2t\bW_0^{\Herm} \circ \bW_1 + t\bW_0 \circ \bW_1^* + t\bW_1^{\Herm}\bW_0 =  t\bW_0^{\Herm}\bW_1 + 2t \bW_1^{\Herm} \circ \bW_0  + t \bW_0^{\Herm} \circ\bW_1 \\
&\implies 2\Im\{\bW_0^{\Herm} \circ \bW_1\}  = \bW_0^{\Herm} \bW_1 - \bW_1^{\Herm}\bW_0 . \label{eq:constraint1implies_simplified}
\end{align}
To simplify the left-hand side of \fref{eq:constraint1implies_simplified}, we use the following:
\begin{align} 
& \bW(t)^{\Herm} \bW(t) = (\bW_0 +t\bW_1)^{\Herm} (\bW_0 +t\bW_1) = \bI +  t\bW_0^{\Herm} \bW_1 +t\bW_1^{\Herm} \bW_0\\
& \implies \frac{d}{dt}(\bW(t)^{\Herm} \bW(t)) = \bW_0^{\Herm} \bW_1 + \bW_1^{\Herm} \bW_0.   \\
& \implies \frac{d}{dt}(\bW(t)^{\Herm} \bW(t))|_{t=0} =\frac{d}{dt}(\bI_N)|_{t=0} = \bZero_{N\times N} \implies \bW_0^{\Herm}\bW_1 + \bW_1^{\Herm}\bW_0  = \bZero_{N\times N}.
\end{align}
Therefore, by inserting $\bW_1^{\Herm}\bW_0  = - \bW_0^{\Herm} \bW_1$ in \fref{eq:constraint1implies_simplified}, we obtain that
\begin{align}
 \Im\{\bW_0^{\Herm} \circ \bW_1\}  = \bW_0^{\Herm} \bW_1. \label{eq:finalW0W1line}
\end{align}
From \fref{eq:finalW0W1line}, we deduce that $\bW_1=\bZero_{N\times N}$ must hold.
As $\bW(t)$ is any arbitrary path, this implies that the variety formed by all the critical points does \textit{not} have a tangent space around $\bW_0$. 
We conclude that the $\bW_0$ is an isolated critical point.

\subsection{Supporting Lemma for \fref{lem:multiple_vecs_conv}}
\begin{lem}\label{lem:single_vec_conv}
Suppose $\vecq$ is a vector on the unit sphere: $\vecq\in \opS^{N-1}$.  For
$\epsilon\in[0,1]$, if $\|\vecq\|_4^4\geq 1-\epsilon$, then  $\exists k\in \setint{N}$  such that
\begin{align}
\vecnorm{\vecq-\vece_k^\theta}_2^2 \leq 2\epsilon,
\end{align}
where $\theta = \angle q_k$, $\vece_k^\theta\in\opC^N$ with $[\vece_k^\theta]_k=e^{j\theta}$, and $[\vece_k^\theta]_\ell=0,\forall \ell\neq k$.
\end{lem}
Let $\vecq = [q_1,q_1,\dots,q_{N}]^{\Tran}$ and without loss of generality, assume
\begin{align} \label{eq:D13}
    1\geq|q_1|^2\geq|q_2|^2\dots\geq|q_{N}|^2\geq 0.
\end{align}
The proof is as follows:
We rewrite the assumption that $\|\vecq\|_4^4\geq 1-\epsilon$ as
\begin{align} \label{eq:D14}
    |q_1|^4+|q_2|^4+\dots+|q_{N}|^4\geq1-\epsilon,
\end{align}
and along with \fref{eq:D13} and $\vecq\in\opS^{N-1}$, we have that
\begin{align}
    |q_1|^4+|q_2|^2(|q_2|^2+\dots+|q_{N}|^2) = |q_1|^4+|q_2|^2(1-|q_1|^2) \geq 1-\epsilon.
\end{align}
Then, as $q_1 \geq q_2$, we have
\begin{align} \label{eq:D16}
|q_1|^4+|q_1|^2(1-|q_1|^2)\geq|q_1|^4+|q_2|^2(1-|q_1|^2) \geq 1-\epsilon \implies |q_1|^2 \geq 1-\epsilon.
\end{align}
Since $\vecq\in\opS^{N-1}$, we then know
\begin{align} \label{eq:D17}
 |q_2|^2+\dots+|q_{N}|^2\leq \epsilon   .
\end{align}
Moreover, as $|q_1|\leq1$, \fref{eq:D16} also implies that
\begin{align} \label{eq:D18}
    \epsilon\geq 1-|q_1|^2 \geq 1-|q_1| \implies (1-|q_1|)^2 \leq \epsilon^2.
\end{align}
Combining \fref{eq:D17} and \fref{eq:D18} with the fact that $\epsilon\leq1$, we have 
\begin{align}
    \vecnorm{|\vecq|-\vece_1^0}_2^2 &= (|q_1|-1)^2+|q_2|^2+\cdots+|q_{N}|^2 \leq \epsilon^2+\epsilon \leq 2\epsilon.
\end{align}
Let $\vece_k^{\theta}=\bme_1^{j\angle q_1}$. Then, we have
$
\vecnorm{\vecq-\vece_k^\theta}_2^2 = \vecnorm{|\vecq|-\vece_1^0}_2^2 \leq 2\epsilon,
$
as desired.
We note that this lemma is an adaptation of \cite [Lemma 36]{zhai20complete} to complex-valued numbers.

\subsection{Supporting Lemma for \fref{lem:closetoCP}}
\begin{lem} \label{lem:multiple_vecs_conv}
Suppose $\vecq_0, \vecq_1,\dots,\vecq_{K-1}$ are $K$ vectors on the unit sphere: $\vecq_k\in\opS^{N-1},\forall k\in[K]$. For $\epsilon\in[0,1]$, if
\begin{align} \label{eq:D20}
    \frac{1}{K}\sum_{k=1}^{K}\vecnorm{\vecq_k}_4^4\geq 1-\epsilon, 
\end{align}
then $\exists m_1,m_2,\dots,m_K\in\setint{N}$ such that 
\begin{align}
    \frac{1}{K}\sum_{k=1}^{K}\vecnorm{\vecq_k -\vece_{m_k}^{\theta_k}} \leq 2\epsilon,
\end{align}
where $\vece_{m_k}^{\theta_k}\in\opC^N$ with $[\vece_{m_k}^{\theta_k}]_{m_k}=e^{j\theta_k}, \theta_k\in\opR$, and $[\vece_{m_k}^{\theta_k}]_{\ell}=0,\forall \ell\neq m_k$.\\
\end{lem}
The proof proceeds as follows:
We can rewrite \fref{eq:D20} as
\begin{align}
    \sum_{k=1}^{K}\vecnorm{\vecq_k}_4^4\geq K-K\epsilon. \label{eq:q4normlowerbound}
\end{align}
Since $\forall k\in[K]$, $0\leq \vecnorm{\vecq_k}_4^4\leq \vecnorm{\vecq_k}_2^4 = 1$, we can assume
\begin{align}
\vecnorm{\vecq_k}_4^4\leq 1-\alpha_k\epsilon, \forall k\in[K],    \label{eq:q4normupperbound}
\end{align}
where we introduce auxiliary variables $\alpha_k$ which satisfy $\alpha_k\epsilon\in[0,1],\forall k\in[K]$  and 
\begin{align}\label{eq:D24}
    \sum_{k=1}^{K}\alpha_k \leq K
\end{align}
must hold to satisfy both \fref{eq:q4normlowerbound} and \fref{eq:q4normupperbound}.
By~\fref{lem:single_vec_conv}, there exists $\vece_{m_k}^{\theta_k}$ such that
\begin{align}
 \vecnorm{\vecq_k - \vece_{m_k}^{\theta_k}}_2^2\leq 2\alpha_k\epsilon, \forall k\in[K] . 
\end{align}
Combined with \fref{eq:D24}, we have
$\frac{1}{K}\sum_{k=1}^{K} \vecnorm{\vecq_k - \vece_{m_k}^{\theta_k}}_2^2 \leq \frac{1}{K}\sum_{k=1}^{K} 2\alpha_k\epsilon = \frac{2\epsilon}{K}\sum_{k=1}^{K}\alpha_k \leq 2\epsilon$, 
as desired.
We note that this lemma is an adaptation of \cite [Lemma 37]{zhai20complete} to complex-valued numbers.

\section{Proofs of~\fref{sec:learning}}

\subsection{Proof of \fref{lem:projU}}
\label{app:projU}

Notice that 
\begin{align} \label{eq:AtoM_F} 
    \vecnorm{\bM-\bA}_{\Fro}^2 &= \tr((\bM-\bA)(\bM-\bA)^{\Herm}) =\tr(\bI-\bA\bM^{\Herm}-\bM\bA^{\Herm}+\bA\bA^{\Herm})\\
    &= N-2\Re\{\tr(\bA\bM^{\Herm})\}+\tr(\bA\bA^{\Herm}).
\end{align}
Since $\tr(\bA\bA^{\Herm})$ is a constant, we know that
\begin{align}
   \underset{\bM\in U(N;\opC)}{\argmin} \vecnorm{\bM-\bA}_{\Fro}^2 = \underset{\bM\in U(N;\opC)}{\argmax} \,\Re\{\tr(\bA\bM^{\Herm})\}.
\end{align}
Let $\bU\bSigma\bV^{\Herm}=\SVD{\bA}$, then
\begin{align}
 \Re\{\tr(\bA\bM^{\Herm})\} &=  \Re\{\tr(\bU\bSigma\bV^{\Herm}\bM^{\Herm})\} =  \Re\{\tr(\bSigma\bV^{\Herm}\bM^{\Herm}\bU)\} \leq \left|\tr(\bSigma\bV^{\Herm}\bM^{\Herm}\bU)\right|\\
 &\leq \sum_{i=1}^{N} \sigma_i(\bA)\sigma_i(\bV^{\Herm}\bM^{\Herm}\bU) = \sum_{i=1}^{N} \sigma_i(\bA),
\end{align}
where the second inequality is obtained through von Neumann's trace inequality,
and the equality holds if and only if $\bV^{\Herm}\bM^{\Herm}\bU$ is a diagonal complex permutation matrix, which implies $\bV^{\Herm}\bM^{\Herm}\bU = \bC\implies\bM=\bU\bC^{\Herm}\bV^{\Herm}$ for some $\bC\in \CP(N), C_{i,k}=0,\forall i\neq k$.
Since \fref{eq:AtoM_F} is invariant to the phase shifts caused by $\bC$, we can take $\bC=\bI$ for simplicity. Hence, $\bM=\bU\bV^{\Herm}$.

\subsection{Proof of \fref{lem:C3}}
\label{app:C3}

Let $\bU\bSigma\bV^{\Herm}=|\bW|^{\circ2}\circ\bW$ denote $\SVD{|\bW|^{\circ2}\circ\bW}$.
\begin{enumerate}[i)]
    \item If $\bW\in U(N; \opC)$ is a fixed point of \fref{alg:msp_yless}, then we have that $ \proj{U(N;\opC)}{|\bW|^{\circ2}\circ\bW}=\bW$;
    equivalently,  $\bW = \bU\bV^{\Herm}$. 
    Using this equality, we simplify $(|\bW|^{\circ2}\circ\bW)^{\Herm} \bW$ as follows:
    \begin{align}  
    (|\bW|^{\circ2}\circ\bW)^{\Herm}\bW  &= (\bU\bSigma\bV^{\Herm})^{\Herm} \bW
       = \bV\bSigma\bU^{\Herm}\bW 
        = \bV\bSigma\bU^{\Herm}\bU\bV^{\Herm}  
        = \bV\bSigma\bV^{\Herm},
    \end{align}
    which implies that $(|\bW|^{\circ2}\circ\bW)^{\Herm}\bW$ is Hermitian symmetric. Hence, we have 
    \begin{align} 
        (|\bW|^{\circ2}\circ\bW)^{\Herm}\bW = \bW^{\Herm}(|\bW|^{\circ2}\circ\bW),
    \end{align}
    and by \fref{lem:C1}, $\bW$ is a critical point of $\ell^4$-norm over the unitary group.
    \item If $\bW$ is a critical point of $\ell^4$-norm over the unitary group, then we have $(|\bW|^{\circ2}\circ\bW)^{\Herm}\bW = \bW^{\Herm}(|\bW|^{\circ2}\circ\bW)$. Then we must have $\SVD{|\bW|^{\circ2}\circ\bW)^{\Herm}\bW} = \SVD{\bW^{\Herm}(|\bW|^{\circ2}\circ\bW)}$. 
    Also, by the rotational invariance of the SVD, we have the following:
    \begin{align}
        &\SVD{(|\bW|^{\circ2}\circ\bW)^{\Herm}\bW} = \bV\bSigma\bU^{\Herm}\bW  =\SVD{\bW^{\Herm}(|\bW|^{\circ2}\circ\bW)} = \bW^{\Herm}\bU\bSigma\bV^{\Herm}.
    \end{align}
   Finally, by the uniqueness of the SVD, we have $\bV=\bW^{\Herm}\bU \implies \bW=\bU\bV^{\Herm}$.
\end{enumerate}

\subsection{Proof of \fref{lem:C4}}
\label{app:C4}

Consider the following objective function $\tilde{h}:U(N;\opC)\to\opR^+ $:
\begin{equation}
    \tilde{h}(\bA) = \begin{cases} \frac{\alpha}{4}\vecnorm{\bA}_4^4+\frac{1}{2}\vecnorm{\bA}_{\Fro}^2 , & \text{when } \alpha <\infty\\
    \vecnorm{\bA}_4^4, &\text{when } \alpha = +\infty.
    \end{cases}
\end{equation}
Note that $\tilde{h}(\bA)$ is convex in both cases. Also note that the Stiefel manifold $U(N;\opC)$ is a compact manifold, so by~\cite[Thm.~1]{JNRS10}, we know that the following iteration update,
\begin{align} \label{eq:C14}
\bA_{k+1} \in \underset{\bW \in U(N;\opC)}{\argmax}\,\, & \tilde{h}(\bA_k)+ \langle \partial \tilde{h}(\bA_k), \bW - \bA_k \rangle 
\end{align}
will find a saddle point of $\tilde{h}(\bA)$ with any initialization $\bA_0\in U(N;\opC)$, where $\langle\cdot,\cdot\rangle: U(N;\opC)\times U(N;\opC) \to \opC$ is defined as
\begin{equation} \label{eq:C15}
    \langle \bW_1 , \bW_2 \rangle = \tr(\bW_1^{\Herm} \bW_2).
\end{equation}
Thus, substituting \fref{eq:C15} into \fref{eq:C14} yields
\begin{equation}
    \bA_{k+1} = \begin{cases} \proj{U(N;\opC)}{\alpha |\bA_k|^{\circ2}\circ\bA_k + \bA_k} &\text{when } \alpha <\infty\\
    \proj{U(N;\opC)}{ |\bA_k|^{\circ2}\circ\bA_k} &\text{when } \alpha = +\infty,
    \end{cases}
\end{equation}
where $\mathcal{P}_{U(N;\opC)}:\opC^{N\times N} \to U(N;\opC)$ is the projection onto $U(N;\opC)$
\begin{align}
    \mathcal{P}_{U(N;\opC)}(\bA) = \bU\bV^{\Herm},\,\, \text{where}\ \bU\bSigma\bV^{\Herm} = \SVD\bA.
\end{align}
Furthermore, notice that $\vecnorm{\bA}_{\Fro}^2=N$ is a constant for unitary $\bA$, thereby finding a critical point of $\tilde{h}(\bA)$ is equivalent to finding a critical point of $\vecnorm{\bA}_4^4$ over $U(N;\opC)$.

\subsection{Proof of \fref{lem:C5}}
\label{app:C5}

Let $\bA = \bD + \bN$ , where $\bD$ is the diagonal part of $\bA$ and $\bN$ is the off-diagonal part. Therefore, we have:
\begin{equation} \label{eq:C18}
 \vecnorm{|\bN|^{\circ2}\circ\bN}_{\Fro} \leq \vecnorm{\bN}_{\Fro}^2 \vecnorm{\bN}_{\Fro} \leq \vecnorm{\bA-\bC}_{\Fro}^{3/2}=\epsilon^{3/2},
\end{equation}
where the first inequality is achieved through Cauchy--Schwarz inequality and the second inequality holds because $\bN$ is the off-diagonal part of $\bA −\bC$.
We can view $|\bA|^{\circ2}\circ\bA$ as $|\bD|^{\circ2}\circ\bD$ plus a small perturbation $|\bN|^{\circ2}\circ\bN$ with norm at most $\epsilon^{3/2}$.
Let $\bU_\bA\bSigma_\bA\bV_\bA^{\Herm} = \SVD{|\bA|^{\circ2}\circ\bA}, \bA'=\bQ_\bA=\bU_\bA\bV_\bA^{\Herm}$ and $\bU_\bD\bSigma_\bD\bV_\bD^{\Herm} = \SVD{|\bD|^{\circ2}\circ\bD}, \bQ_\bD=\bU_\bD\bV_\bD^{\Herm}$.
By~\cite[Thm.~1]{Li95}, we have
\begin{align} \label{eq:C19}
    \vecnorm{\bQ_\bA-\bQ_\bD}_{\Fro} &\leq \frac{2\vecnorm{|\bA|^{\circ2}\circ\bA - |\bD|^{\circ2}\circ\bD}_{\Fro}}{\sigma_N(|\bA|^{\circ2}\circ\bA) + \sigma_N(|\bD|^{\circ2}\circ\bD)} = \frac{2\vecnorm{|\bN|^{\circ2}\circ\bN}_{\Fro}}{\sigma_N(|\bA|^{\circ2}\circ\bA) + \sigma_N(|\bD|^{\circ2}\circ\bD)}.
\end{align}
Since $|\bD|^{\circ2}\circ\bD$ is a diagonal matrix, $\bQ_D=\bI_N$, and $\sigma_N(|\bD|^{\circ2}\circ\bD)=\min_k |A_{k,k}|^3$. Furthermore,
\begin{align}
\epsilon &= \vecnorm{\bA-\bC}_{\Fro}^2 = \sum_{k=1}^{N} (|A_{k,k}|-1)^2 + \sum_{k=1}^{N} \sum_{\substack{\ell=1 \\ \ell\neq k}}^{N} |A_{k,\ell}|^2 
= \sum_{k=1}^{N}\sum_{\ell=1}^{N} |A_{k,\ell}|^2 - 2\sum_{k=1}^{N} |A_{k,k}| + N \\
&= 2N- 2\sum_{k=1}^{N} |A_{k,k}| 
    \iff\sum_{k=1}^{N} A_{k,k} = N-\frac{\epsilon}{2}.
\end{align}
Without loss of generality, we can assume $1\geq|A_{0,0}|\geq|A_{1,1}|\geq\dots\geq|A_{N-1,N-1}|>0$, so we have
\begin{equation}
    |A_{N-1,N-1}| = N-\frac{\epsilon}{2} - \sum_{k=1}^{N-1}|A_{k,k}| \geq N-\frac{\epsilon}{2}-(N-1)=1-\frac{\epsilon}{2},
\end{equation}
which implies that
\begin{equation} \label{eq:C22}
\sigma_N(|\bD|^{\circ2}\circ\bD)=\min_k |A_{k,k}|^3 \geq \left(1-\frac{\epsilon}{2}\right)^3 .
\end{equation}
We use \cite[Thm.~2.17]{DV92} as stated in \cite[Lem.~46]{zhai20complete} as follows:
 Let $\bG \in\opC^{N\times N}$ be a general full rank matrix and $\bG= \bB\bM$, where $\bM$ is a diagonal matrix with $M_{k,k}= \|\bmg_k\|_2$ so that $\bB$ has unit matrix-two norm, i.e., $\|\bB\|_2=1$. %
Let $\delta \bG = \delta \bB\bM$ be a perturbation of $\bG$ such that $\|\delta \bB\|_2 \leq \sigma_{N}(\bB)$.
Then, $\frac{\|\sigma_i(\bG) - \sigma_i(\delta\bG)|}{\sigma_i(\bG)} \leq \frac{\|\delta\bB\|_2}{\sigma_N(\bB)}$.

We substitute the variables as $\bG = |\bD|^{\circ2}\circ\bD = \bB\bM$ and $\delta\bG =  |\bA|^{\circ2}\circ\bA -  |\bD|^{\circ2}\circ\bD =  |\bN|^{\circ2}\circ\bN = \delta\bB\bM$.
Here, we know $\bB=\bI,\bM=|\bD|^{\circ2}\circ\bD, \delta\bB= \delta\bG \bM^{-1} = (|\bN|^{\circ2}\circ\bN)(|\bD|^{\circ2}\circ\bD)^{-1}$, and thus
\begin{align}\label{eq:C24}
    \frac{|\sigma_N(|\bA|^{\circ2}\circ\bA)-\sigma_N(|\bD|^{\circ2}\circ\bD)|}{\sigma_N(|\bD|^{\circ2}\circ\bD)} \leq \frac{\vecnorm{\delta\bB}_2}{\sigma_N(\bB)} = \vecnorm{\delta\bB}_2,
\end{align}
which implies 
\begin{align}
&|\sigma_N(|\bA|^{\circ2}\circ\bA)-\sigma_N(|\bD|^{\circ2}\circ\bD)| \leq \vecnorm{(|\bN|^{\circ2}\circ\bN)(|\bD|^{\circ2}\circ\bD)^{-1}}_2 \sigma_N(|\bD|^{\circ2}\circ\bD) \\
&\leq \vecnorm{|\bN|^{\circ2}\circ\bN}_2 \vecnorm{(|\bD|^{\circ2}\circ\bD)^{-1}}_2 \sigma_N(|\bD|^{\circ2}\circ\bD) \\
&\leq \vecnorm{|\bN|^{\circ2}\circ\bN}_2 \left(1-\frac{\epsilon}{2}\right)^{-3} \sigma_N(|\bD|^{\circ2}\circ\bD)\\
&\leq \epsilon^{3/2} \left(1-\frac{\epsilon}{2}\right)^{-3}  \sigma_N(|\bD|^{\circ2}\circ\bD), \label{eq:C25} 
\end{align}
where the four inequalities above follow from \fref{eq:C24}, Cauchy--Schwarz inequality, \fref{eq:C22}, and \fref{eq:C18}, respectively. 
Then, we have 
\begin{align} \label{eq:C25_2} 
- \epsilon^{3/2} \bigg(1-\frac{\epsilon}{2}\bigg)^{-3}  \sigma_N(|\bD|^{\circ2}\circ\bD) &\leq \sigma_N(|\bA|^{\circ2}\circ\bA)-\sigma_N(|\bD|^{\circ2}\circ\bD) .
\end{align}
Therefore, we have the following two inequalities  from \fref{eq:C25_2} and \fref{eq:C22}, respectively:
\begin{align}  
\sigma_N (|\bA|^{\circ2}\circ\bA)+\sigma_N(|\bD|^{\circ2}\circ\bD) &\geq \left( 2- \epsilon^{3/2} \left(1-\frac{\epsilon}{2}\right)^{-3} \right) \sigma_N(|\bD|^{\circ2}\circ\bD) \\
&\geq 2\left(1-\frac{\epsilon}{2}\right)^3 - \epsilon^{3/2} . \label{eq:C26}
\end{align}
By using \fref{eq:C18} and \fref{eq:C26} in \fref{eq:C19}, we obtain 
\begin{align} 
    \vecnorm{\bQ_\bA-\bQ_\bD}_{\Fro} &\leq \frac{2\vecnorm{|\bN|^{\circ2}\circ\bN}_{\Fro}}{\sigma_N(|\bA|^{\circ2}\circ\bA) + \sigma_N(|\bD|^{\circ2}\circ\bD)} \leq \frac{2\epsilon^{3/2}}{2\left(1-\frac{\epsilon}{2}\right)^3-\epsilon^{3/2}}.
\end{align}
Recalling that $\bQ_\bA = \bA'$ and $\bQ_D = \bI_N$, we have
$\vecnorm{\bA'-\bC'}_{\Fro} \leq \vecnorm{\bA'-\bI}_{\Fro} = \vecnorm{\bQ_\bA-\bQ_\bD}_{\Fro} \leq O(\epsilon^3)$.
Moreover, such operation is a contraction whenever 
\begin{align}
&\frac{2\epsilon^{3/2}}{2\left(1-\frac{\epsilon}{2}\right)^3-\epsilon^{3/2}} \leq \vecnorm{\bA-\bD}_{\Fro} = \epsilon^{1/2}  \\
&\iff 2\epsilon -2 \left(1-\frac{\epsilon}{2}\right)^3  + \epsilon^{3/2} \leq 0 \iff \epsilon <  0.394,
\end{align} 
which completes the proof. 

\subsection{Proof of \fref{lem:UVH}}
\label{app:UVH}

Since the left and right singular vectors of full rank matrices are unique up to complex phase {shifts}, $\bU\bV^{\Herm}$ is unique. Since $\bA$ is unitary, one SVD of $\bA$ is $\bU=\bI_N$, $\bSigma = \bI_N$, $\bV=\bA$.
Then $\bA_1=\bD\bA=\bD\bV=\bU_1\bSigma_1\bV_1$, so there exists an SVD of $\bA_1$ such that $\bU_1=\bI_N, \bSigma_1=\bD$, and $\bV_1 = \bA$. Another SVD of $\bA$ is $\bU=\bA$, $\bSigma = \bI_N$, and $\bV=\bI_N$.
Then, $\bA_2=\bA\bD=\bU\bD=\bU_2\bSigma_2\bV_2$, so there exists an SVD of $\bA_2$ such that $\bU_2=\bA, \bSigma_2=\bD$, and $\bV_2 = \bI_N$. Consequently, we have $\bU_1\bV_1^{\Herm}=\bU_2\bV_2^{\Herm} = \bA$.

\subsection{Proof of \fref{lem:fpsofMSP}}
\label{app:fpsofMSP}

Let $\bA_t = \bA$. If $\nabla_{\bA_t} \gint{\bA_t } = \bD\bA_t$ or $\nabla_{\bA_t} \gint{\bA_t } = \bA_t\bD$ is satisfied,
then, by~\fref{lem:UVH},
$ \bA_{t+1} = \proj{U(N;\opC)}{\nabla_{\bA_t} \gint{\bA_t }} = \bA_t$ holds. 
Hence, the matrix $\bA$ is a fixed point of~\fref{alg:msp_int}.

\section{Proofs of~\fref{sec:optimality}}

\subsection{Proof of~\fref{thm:dft_msp_fp}}
\label{app:dft_msp_fp}

Let $\bF=\bF_B$ for simplicity. 
We first expand the objective function in~\fref{eq:gint} as follows:
\begin{equation}
    \gint{\bA} =  \int_{\omega_1} \cdots \int_{\omega_L }\vecnorm{\bA\vecy }_4^4 f_{\Omega_1}(\omega_1) \cdots f_{\Omega_L}(\omega_L) \td\omega_1 \cdots \td\omega_L. \label{eq:multipathgint}
\end{equation}
By setting $\bA = \bF$ and inserting $f_\Omega(\omega_\ell)=\frac{1}{2\pi} \mathbbm{1}\{\omega\in (0,2\pi)\}$ in \fref{eq:multipathgint}, we obtain the following objective function:
\begin{align}
&\gint{\bF  } = \frac{1}{(2\pi)^L} \int_{\omega_1=0}^{2\pi} \cdots \int_{\omega_L=0}^{2\pi}\vecnorm{\bF\vecy}_4^4 \td\omega_1 \cdots \td\omega_L\\
&=   \frac{1}{(2\pi)^L} \int_{\omega_1=0}^{2\pi} \cdots \int_{\omega_L=0}^{2\pi} \sum_{i=1}^B \bigg|{\sum_{n=1}^B F_{i,n} y_n }\bigg|^4 \td\omega_1 \cdots \td\omega_L \\
&= \sum_{i=1}^B   \sum_{n=1}^B \sum_{p=1}^B \sum_{q=1}^B \sum_{r=1}^B F_{i,n}^*  F_{i,p} F_{i,q}^*  F_{i,r}  C_L(p,q,r,n), \label{eq:g_F_fivesums_multipath}
\end{align}
where we define $C_L(p,q,r,n) \define \frac{1}{(2\pi)^L}   \int_{\omega_1=0}^{2\pi} \dots \int_{\omega_L=0}^{2\pi}  { y_p } { y_q^* } {y_r}{y_n^*} \td\omega_1 \cdots \td\omega_L$.
Inserting the expansion of $y_n$ from~\fref{eq:ejw}, we have that
\begin{align}
C_L(p,q,r,n)  =  \frac{1}{(2\pi)^L}   \int_{\omega_1=0}^{2\pi} \dots \int_{\omega_L=0}^{2\pi} &\bigg(\sum_{\ell=1}^L c_\ell e^{j\omega_\ell p}\bigg) \bigg(\sum_{\ell=1}^L c_\ell^* e^{-j\omega_\ell q}\bigg) \bigg(\sum_{\ell=1}^L c_\ell e^{j\omega_\ell r}\bigg) 
\times\! \bigg(\sum_{\ell=1}^L c_\ell^* e^{-j\omega_\ell n}\bigg) \td\omega_1\dots \td\omega_L. \label{eq:Cpmnk}
\end{align}
To establish that $\bF$ is a fixed point, we utilize~\fref{lem:fpsofMSP} to show that 
there exists a full-rank diagonal matrix $\bD\in\opC^{B\times B}$ such that $\nabla_\bF \gint{\bF }= \bF\bD$, i.e.,
\begin{equation}\label{eq:aim_dftopt}
\pder{\gint{\bF  }}{\vecf_n^*} = D_{n,n}\vecf_n,\, \forall n\in \setint{B}.
\end{equation}
Equivalently, for all pairs of indices $(i,n)$, we need to show that
\begin{equation} \label{eq:aim_equiv_eqDkk_multipath}
\pder{\gint{\bF }}{F_{i,n}^*} F_{i,n}^*= D_{n,n},
\end{equation}
which implies that  $\pder{\gint{\bF }}{F_{i,n}^*} F_{i,n}^*$ depends only on $n$ (and does not depend on $i$) for all pair of indices $(i,n)$.
By inserting $F_{i,p} = \frac{1}{\sqrt{B}}  e^{\frac{-j 2\pi (i-1) (p-1)}{B}}$ into~\fref{eq:g_F_fivesums_multipath}, we calculate the gradient as follows:
\begin{align}  
&\pder{\gint{\bF }}{F_{i,n}^*} =  \sum_{p=1}^{B}\sum_{q=1}^{B}\sum_{r=1}^{B} F_{i,p} F_{i,q}^* F_{i,r} C_L(p,q,r,n) =
\frac{1}{B\sqrt{B}}  \sum_{p=1}^{B}\sum_{q=1}^{B}\sum_{r=1}^{B} e^{\frac{-j 2\pi}{B} (i-1) ((p-1)-(q-1)+(r-1))}  C_L(p,q,r,n).
\end{align}
Simplifying the gradient expression with a change in the subscripts, we obtain that
\begin{align} \label{eq:gradFNEW_multipath}
\pder{\gint{\bF }}{F_{i+1,n+1}^*} 
&=  \frac{1}{B\sqrt{B}} \sum_{p=0}^{B-1}\sum_{q=0}^{B-1}\sum_{r=0}^{B-1}  e^{\frac{-j 2\pi i (p-q+r)}{B}} C_L(p,q,r,n).
\end{align}
Multiplying both sides of~\fref{eq:gradFNEW_multipath} by $F_{i+1,n+1}^* =  \frac{1}{\sqrt{B}}e^{\frac{j 2\pi i n}{B}}$, we obtain that
\begin{align}
\pder{\gint{\bF }}{F_{i+1,n+1}^*}  F_{i+1,n+1}^* = \frac{1}{B^2} \sum_{p=0}^{B-1}\sum_{q=0}^{B-1}\sum_{r=0}^{B-1}  e^{\frac{-j 2\pi i(p-q+r-n)}{B}}C_L(p,q,r,n). \label{eq:aim_st_multipath}
\end{align}
Now, let us examine $C_L(p,q,r,n)$: For $L=1$, we have that
\begin{align}
 C_1(p,q,r,n) &= \frac{|c_1|^4}{2\pi} \int_{0}^{2\pi}  e^{j\omega (p-q+r-n)} \td\omega =  |c_1|^4  \delta[p-q+r-n]. \label{eq:C_L1}
\end{align}
For $L=2$, we have that 
\begin{align}
&C_2(p,q,r,n) 
= \frac{1}{(2\pi)^2} \int_{\omega_1=0}^{2\pi} \int_{\omega_2=0}^{2\pi} \bigg(\sum_{\ell=1}^2 c_\ell e^{j\omega_\ell p}\bigg)\! \bigg(\sum_{\ell=1}^2 c_\ell^* e^{-j\omega_\ell q}\bigg)\! \bigg(\sum_{\ell=1}^2 c_\ell e^{j\omega_\ell r}\bigg) \! \bigg(\sum_{\ell=1}^2 c_\ell^* e^{-j\omega_\ell n}\bigg) \td{\omega_1}\td{\omega_2} \\
&=  (|c_1|^4 + |c_2|^4) \delta[p-q+r-n] \notag   + |c_1|^2 |c_2|^2 \delta[p-q]\delta[r-n]  + |c_1|^2c_1 c_2^* \delta[p-q+r]\delta[n]  \notag\\
&+ |c_1|^2c_1^* c_2 \delta[p-q-n]\delta[r] + |c_1|^2 |c_2|^2 \delta[p-q]\delta[r-n]  + c_1 |c_2|^2 c_2^* \delta[r]\delta[p-q-n] \notag \\
&  + c_1^* |c_2|^2 c_2 \delta[n] \delta[p-q+r] + |c_1|^2 c_1 c_2^*  \delta[p+r-n] \delta[q] + c_1 |c_2|^2 c_2^*\delta[p] \delta[-q+r-n] \notag  \\
&  + c_1^2 (c_2^*)^2 \delta[p+r]\delta[q+n] + |c_1|^2|c_2|^2 \delta[p-n]\delta[-q+r] + |c_1|^2 c_1^* c_2 \delta[-q+r-n]\delta[p]  \notag \\ 
&  + c_1^* |c_2|^2c_2\delta[q] \delta[p+r-n] +  |c_1|^2 |c_2|^2 \delta[-q+r]\delta[p-n] + (c_1^*)^2 c_2^2\delta[q+n]\delta[p+r] .
\label{eq:C_L2}
 \end{align} 
We observe the following: 
For $L=1$, $C_L(p,q,r,n)$ is equal to a scaled $\delta[p-q+r-n]$. 
For $L=2$, $C_L(p,q,r,n)$ is a linear combination of (i) $\delta[p-q+r-n]$ terms  and (ii) terms which are a multiplication of two $\delta[\cdot]$-functions; in both cases, each term is nonzero if and only if  $p-q+r-n=0$.
Similarly, for any $L$, $C_L(p,q,r,n)$ is a linear combination of terms which are a multiplication of up to $L$ $\delta[\cdot]$-functions, where each term is nonzero if and only if  $p-q+r-n=0$ holds.
Hence, we rewrite   \fref{eq:aim_st_multipath} by setting $p-q+r-n=0$  as 
\begin{align}
\pder{\gint{\bF }}{F_{i+1,n+1}^*}  F_{i+1,n+1}^* = \frac{1}{B^2} \sum_{p=0}^{B-1}\sum_{q=0}^{B-1}\sum_{r=0}^{B-1}  C_L(p,q,r,n).
\end{align}
Clearly, $\pder{\gint{\bF }}{F_{i,n}^*}  F_{i,n}^* $ does not depend on $i$, 
which yields that $\bF$ is a fixed point of~\fref{alg:msp_int} for the stochastic data model $\vecy $ as given in~\fref{eq:ejw} for any $L$.

\subsection{Proof of~\fref{thm:dft_opt}}
\label{app:dft_opt}

In the following, we retain $L$ as a variable throughout and  impose $L=1$ only in the final step to highlight where this constraint becomes necessary. 

Our goal is to show that the first- and second-derivative conditions from \fref{lem:ca_opt} are satisfied.
By rewriting the first-derivative expression in~\fref{eq:first_deriv} for $\bmy$ as in~\fref{eq:ejw}, setting $\bA = \bF$ and inserting $f_\Omega(\omega_\ell)=\frac{1}{2\pi} \mathbbm{1}\{\omega\in (0,2\pi)\}$, our goal is to show that the following equation holds:
\begin{align}
& \pder{ \int_{\omega_1} \cdots \int_{\omega_L}   \vecnorm{\bG(i,k,\alpha) \vecx }_4^4 f_{\Omega_1}(\omega_1) \cdots f_{\Omega_L}(\omega_L) \td\omega_1 \cdots \td\omega_L }{\alpha}\bigg|_{\alpha=0} \notag \\
&= \frac{4}{(2\pi)^L} \int_{\omega_1=0}^{2\pi} \dots \int_{\omega_L=0}^{2\pi}   \Re\{ x_ix_k(x_i^*)^2 - x_i x_k (x_k^*)^2  \}  \td\omega_1 \cdots \td\omega_L = 0, \label{eq:firstderivmultipath}
\end{align}
where $\vecx  = \bF \vecy  $.
For showing that~\fref{eq:firstderivmultipath} holds, we first simplify the following integral:
\begin{align}
& \frac{4}{(2\pi)^L}\int_{\omega_1=0}^{2\pi} \dots \int_{\omega_L=0}^{2\pi}   x_i x_k (x_i^*)^2 \td\omega_1\dots \td\omega_L \notag \\
 &= \frac{4}{(2\pi)^L}\sum_{p=1}^{B}\sum_{q=1}^{B}\sum_{r=1}^{B} \sum_{n=1}^{B} F_{i,p} F_{k,r}  F_{i,q}^* F_{i,n}^*   \int_{\omega_1=0}^{2\pi} \dots \int_{\omega_L=0}^{2\pi}  y_p y_q^* y_r  y_n^*  \td\omega_1\dots \td\omega_L  \\
 &=  \frac{4}{B^2}\sum_{p=0}^{B-1}\sum_{q=0}^{B-1}\sum_{r=0}^{B-1} \sum_{n=0}^{B-1}  e^{-\frac{j 2\pi (kr + i(p-q-n))}{B} }  C_L(p,q,r,n). \label{eq:scaledCpqrn}
\end{align}
Here, $ C_L(p,q,r,n)$ follows the definition from~\fref{eq:Cpmnk}. 
From the proof of~\fref{thm:dft_msp_fp}, we know that $C_L(p,q,r,n)$ is nonzero if and only if $p-q+r-n=0$ holds.
Therefore, 
 \begin{align}
& \frac{4}{(2\pi)^L}\int_{\omega_1=0}^{2\pi} \dots \int_{\omega_L=0}^{2\pi}   x_i x_k (x_i^*)^2 \td\omega_1\dots \td\omega_L 
 =  \frac{4}{B^2}\sum_{p=0}^{B-1}\sum_{q=0}^{B-1}\sum_{r=0}^{B-1} \sum_{n=0}^{B-1}  e^{\frac{j 2\pi (i-k)r}{B} }  C_L(p,q,r,n) . \label{eq:forsumeCL}
\end{align}
Observing the dependence of the left-hand side of~\fref{eq:forsumeCL} on $(i-k)$, we deduce that
\begin{align}
\int_{\omega_1=0}^{2\pi} \dots \int_{\omega_L=0}^{2\pi}   x_i x_k (x_i^*)^2 \td\omega_1\dots \td\omega_L = \bigg( \int_{\omega_1=0}^{2\pi} \dots \int_{\omega_L=0}^{2\pi}   x_i x_k (x_k^*)^2 \td\omega_1\dots \td\omega_L \bigg)^*,    
\end{align}
which implies that \fref{eq:firstderivmultipath} holds. 
Therefore, for any $L$, we have that the function $\vecnorm{\bG(i,k,\alpha_{i,k})\bF\vecy}_4^4$ from~\fref{eq:ca_alpha} attains a stationary point at $\alpha_{i,k}=0$ for all $ i,k\in \setint{N} ,\, i>k$;
with this, we have proven that the first condition from~\fref{lem:ca_opt} holds.

By rewriting the second-derivative expression in~\fref{eq:second_deriv} for $\bmy$ as in~\fref{eq:ejw}, setting $\bA = \bF$ and inserting $f_\Omega(\omega_\ell)=\frac{1}{2\pi} \mathbbm{1}\{\omega\in (0,2\pi)\}$, we define 
\begin{align}
&d_L(i,k) \define \frac{4}{(2\pi)^L} \int_{\omega_1=0}^{2\pi} \dots \int_{\omega_L=0}^{2\pi}   \big( 2\Re\{x_k^2(x_i^*)^2\} + 4|x_i|^2|x_k|^2 - |x_k|^4 - |x_i|^4 \big)   \td\omega_1 \cdots \td\omega_L . %
\end{align}
Our goal is to show that $ d_L(i,k)< 0$ holds.
In the following steps, we again use that $C_L(p,q,r,n)$ is nonzero if and only if $p-q+r-n=0$ holds.
Let us first consider the integral of the first term with $2\Re\{x_k^2(x_i^*)^2\}$ by initially simplifying the integral of $x_k^2(x_i^*)^2$:
\begin{align}
&\frac{4}{(2\pi)^L} \int_{\omega_1=0}^{2\pi} \dots \int_{\omega_L=0}^{2\pi}   x_k^2(x_i^*)^2   \td\omega_1 \cdots \td\omega_L \notag \\ 
&= \frac{4}{(2\pi)^L}\sum_{p=1}^{B}\sum_{q=1}^{B}\sum_{r=1}^{B} \sum_{n=1}^{B} F_{k,p} F_{k,r}  F_{i,q}^* F_{i,n}^*   \int_{\omega_1=0}^{2\pi} \dots \int_{\omega_L=0}^{2\pi}  y_p y_q^* y_r  y_n^*  \td\omega_1\dots \td\omega_L  \\
 &=  \frac{4}{B^2}\sum_{p=0}^{B-1}\sum_{q=0}^{B-1}\sum_{r=0}^{B-1} \sum_{n=0}^{B-1}  e^{-\frac{j 2\pi (k(p+r) + i(-q-n))}{B} }  C_L(p,q,r,n)\\
 &= \frac{4}{B^2}\sum_{p=0}^{B-1}\sum_{q=0}^{B-1}\sum_{r=0}^{B-1} \sum_{n=0}^{B-1}  e^{\frac{j 2\pi (i-k)(p+r)}{B} }  C_L(p,q,r,n)
\end{align}
Consequently, we have that
\begin{align}
&\frac{4}{(2\pi)^L} \int_{\omega_1=0}^{2\pi} \dots \int_{\omega_L=0}^{2\pi}   \big( 2\Re\{x_k^2(x_i^*)^2\}  \big)   \td\omega_1 \cdots \td\omega_L 
 \notag \\&
= \frac{8}{B^2}\sum_{p=0}^{B-1}\sum_{q=0}^{B-1}\sum_{r=0}^{B-1} \sum_{n=0}^{B-1}   \cos\bigg({\frac{2\pi}{B} (i-k)(p+r) }\bigg)  \Re\{C_L(p,q,r,n)\}.
\end{align}
Now, we simplify the integral for the second term:
\begin{align}
&\frac{16}{(2\pi)^L} \int_{\omega_1=0}^{2\pi} \dots \int_{\omega_L=0}^{2\pi} x_i x_i^* x_k x_k^*    \td\omega_1 \cdots \td\omega_L \notag\\ 
&= \frac{16}{(2\pi)^L}\sum_{p=1}^{B}\sum_{q=1}^{B}\sum_{r=1}^{B} \sum_{n=1}^{B} F_{i,p} F_{k,r}  F_{i,q}^* F_{k,n}^*   \int_{\omega_1=0}^{2\pi} \dots \int_{\omega_L=0}^{2\pi}  y_p y_q^* y_r  y_n^*  \td\omega_1\dots \td\omega_L  \\
 &=  \frac{16}{B^2}\sum_{p=0}^{B-1}\sum_{q=0}^{B-1}\sum_{r=0}^{B-1} \sum_{n=0}^{B-1}  e^{-\frac{j 2\pi (k(r-n) + i(p-q))}{B} }  C_L(p,q,r,n)\\ 
 &= \frac{16}{B^2}\sum_{p=0}^{B-1}\sum_{q=0}^{B-1}\sum_{r=0}^{B-1} \sum_{n=0}^{B-1}  e^{\frac{j 2\pi (i-k)(r-n)}{B} }  C_L(p,q,r,n)\\
 &= \frac{16}{B^2}\sum_{p=0}^{B-1}\sum_{q=0}^{B-1}\sum_{r=0}^{B-1} \sum_{n=0}^{B-1} \cos\bigg({\frac{2\pi}{B} (i-k)(r-n) }\bigg)   \Re\{C_L(p,q,r,n)\},
\end{align}
where the last step follows from knowing that the integral has to be real-valued and nonnegative as the integration is over absolute squares. 
Next, we simplify the integral for the third term:
\begin{align}
&\frac{4}{(2\pi)^L} \int_{\omega_1=0}^{2\pi} \dots \int_{\omega_L=0}^{2\pi} x_i x_i^* x_i x_i^*    \td\omega_1 \cdots \td\omega_L \\ 
&= \frac{4}{(2\pi)^L}\sum_{p=1}^{B}\sum_{q=1}^{B}\sum_{r=1}^{B} \sum_{n=1}^{B} F_{i,p} F_{i,r}  F_{i,q}^* F_{i,n}^*   \int_{\omega_1=0}^{2\pi} \dots \int_{\omega_L=0}^{2\pi}  y_p y_q^* y_r  y_n^*  \td\omega_1\dots \td\omega_L  \\
 &=  \frac{4}{B^2}\sum_{p=0}^{B-1}\sum_{q=0}^{B-1}\sum_{r=0}^{B-1} \sum_{n=0}^{B-1}  e^{-\frac{j 2\pi i(p-q+r-n)}{B} }  C_L(p,q,r,n)\\
 &= \frac{4}{B^2}\sum_{p=0}^{B-1}\sum_{q=0}^{B-1}\sum_{r=0}^{B-1} \sum_{n=0}^{B-1}   C_L(p,q,r,n) = \frac{4}{B^2}\sum_{p=0}^{B-1}\sum_{q=0}^{B-1}\sum_{r=0}^{B-1} \sum_{n=0}^{B-1}  \Re\{ C_L(p,q,r,n) \}, \label{eq:simplifiedthirdterm2ndderiv}
\end{align}
which evidently does not depend on $i$; therefore, we deduce that the integral for the fourth term is equal to that of the third.
Finally, we have that
\begin{align}
d_L(i,k) = \frac{4}{B^2}\sum_{p=0}^{B-1}\sum_{q=0}^{B-1}\sum_{r=0}^{B-1} \sum_{n=0}^{B-1}  \bigg( &2  \cos\bigg({\frac{2\pi}{B} (i-k)(p+r) }\bigg)
+ 4\cos\bigg({\frac{2\pi}{B} (i-k)(r-n)}\bigg) -2 \bigg)  \Re\{C_L(p,q,r,n)\} .\label{eq:dft_ca_2nd_deriv}
\end{align}
Unfortunately, we are unable to show that $d_L(i,k)<0$ holds for a multipath model with any $L$; therefore, we cannot claim whether the DFT attains a local maximum for any $L$.
However, for a single-path model with $L=1$, we are able to simplify $d_L(i,k)$ as 
$d_1(i,k) = \frac{8}{B^2} \left(3B\csc^2(\pi(i-k)/B) - \frac{2B^3+7B}{3}\right)$.
Now, we aim to show the following:
\begin{align}
d_1(i,k)<0 \Leftrightarrow
\csc^2(\pi(i-k)/B) < \frac{2B^2+7}{9} 
\Leftrightarrow\sin(\pi(i-k)/B) > \frac{3}{\sqrt{2B^2+7}} . 
\end{align}
For $B\geq 2 $, we have that
\begin{align}
\sin(\pi (i-k)/B) \geq \sin(\pi /B) \geq \frac{\pi}{B}-\frac{1}{6}\bigg(\frac{\pi}{B}\bigg)^3 > \frac{3}{\sqrt{2B^2+7}}, \label{eq:finalinequalitiesfor2ndderiv}
\end{align}
where the first inequality holds because $ (i-k)\in \{1,\dots,B-1\} $,
the second inequality follows from the lower bound in~\cite[Theorem  3.1]{sinLB}, and
the last inequality is satisfied for $B\geq 1.75$.
Hence, we have proven with \fref{eq:finalinequalitiesfor2ndderiv} that $d_1(i,k)<0,\forall i,k\in\setint{B}, i>k $ holds, 
which implies that that the second condition from~\fref{lem:ca_opt} holds and concludes our proof.

\subsection{Simplified Gradient  for the DCT}
\label{app:dct_subopt}

Let $\vecx \define \bC_B\vecy $. We have that
\begin{align} \label{eq:xk_cos}
x_{k+1} &= \sqrt{\frac{2}{B}}  \frac{1}{\sqrt{1+\delta[k]}} \sum_{n=0}^{B-1} \cos(\omega n + \phi) \cos\left(\frac{\pi}{B}(n+0.5)k\right) .    
\end{align}
Since $\vecx \in\opR^B$,  $f_\Omega(\omega_\ell)=\frac{1}{2\pi} \mathbbm{1}\{\omega\in (0,2\pi)\}$, and $f_\Phi(\phi) = \frac{1}{2\pi} \mathbbm{1}\{\phi\in (0,2\pi)\}$, we rewrite~\fref{eq:first_deriv} as
\begin{align} \label{eq:real_first_deriv}
\pder{\int_0^{2\pi}\int_0^{2\pi}\vecnorm{\bG(i,k,\alpha)\vecx }_4^4 f_\Omega(\omega) f_\Phi(\phi)\td\omega \td\phi }{\alpha}\bigg|_{\alpha=0}  
& = \frac{4}{(2\pi)^2}\int_0^{2\pi}\left( x_i^3x_k - x_k^3x_i \right) \td{\omega} .
\end{align}
Inserting $x_k$ from~\fref{eq:xk_cos} into~\fref{eq:first_deriv} followed by a sequence of tedious algebraic simplifications with trigonometric identities yield that 
\begin{align}   
&\pder{\int_0^{2\pi}\int_0^{2\pi}\vecnorm{\bG(i,k,\alpha)\vecx }_4^4 f_\Omega(\omega) f_\Phi(\phi)\td\omega \td\phi }{\alpha}\bigg|_{\alpha=0}\notag \\ 
&=  
\frac{1}{B^2\sqrt{1+\delta[k]}}  
\sum_{p=0}^{B-1} \sum_{q=0}^{B-1} \sum_{r=0}^{B-1} \sum_{n=0}^{B-1}  \cos\left(\frac{\pi}{B}\bigg(p+\half\bigg)k\right)\cos\left(\frac{\pi}{B}\bigg(q+\half\bigg)i\right) \notag\\
&\quad \times 
\left(  
\cos\left(\frac{\pi}{B}\bigg(r+\half\bigg)i\right)\cos\left(\frac{\pi}{B}\bigg(n+\half\bigg)i\right) - \frac{1}{1+\delta[k]}\cos\left(\frac{\pi}{B}\bigg(r+\half\bigg)k\right)\cos\left(\frac{\pi}{B}\bigg(n+\half\bigg)k\right)
\right) 
\notag \\
&\quad \times \big(  \delta[p-q+r-n] + \delta[p-q+r-n] + \delta[p-q-r+n] \big).
\end{align}

\bibliographystyle{IEEEtran}

\bibliography{bib/IEEEabrv,bib/confs-jrnls,bib/publishers,bibfile,bib/vipbib}

\end{document}